\documentclass[twocolumn,twocolappendix,10pt]{aastex631}
\usepackage{placeins,verbatim}
\usepackage{xcolor}

\newcommand{\del}[1]{}

\renewcommand{\edit}[1]{}

\sfcode`A=1000 \sfcode`B=1000 \sfcode`C=1000 \sfcode`D=1000
\sfcode`E=1000 \sfcode`F=1000 \sfcode`G=1000 \sfcode`H=1000
\sfcode`I=1000 \sfcode`J=1000 \sfcode`K=1000 \sfcode`L=1000
\sfcode`M=1000 \sfcode`N=1000 \sfcode`O=1000 \sfcode`P=1000
\sfcode`Q=1000 \sfcode`R=1000 \sfcode`S=1000 \sfcode`T=1000
\sfcode`U=1000 \sfcode`V=1000 \sfcode`W=1000 \sfcode`X=1000
\sfcode`Y=1000 \sfcode`Z=1000

\newcommand{\Msol}{\hbox{M$_{\odot}$}}
\newcommand{\hii}{\hbox{\ion{H}{2}}}
\newcommand{\no}{\nodata}
\newcommand{\eg}{e.g.}
\newcommand{\ie}{i.e.}
\newcommand{\mmJy}{\hbox{$\mu$Jy}}
\newcommand{\zp}{\ensuremath{z_{\rm ph}}}
\newcommand{\zph}{\ensuremath{z_{\rm ph}}}
\newcommand{\zsp}{\ensuremath{z_{\rm sp}}}
\newcommand{\etal}{et~al.}

\begin{document}

\title{PEARLS: JWST counterparts of micro-Jy radio sources in the Time Domain Field}
\shorttitle{Radio Counterparts in the TDF}
\shortauthors{Willner, Gim, Poletta \etal}

\author[0000-0002-9895-5758]{S.\ P.\ Willner}
\affiliation{Center for Astrophysics \textbar\ Harvard \& Smithsonian, 60 Garden Street, Cambridge, MA, 02138, USA}

\correspondingauthor{S.\ P.\ Willner}
\email{swillner@cfa.harvard.edu}

\author[0000-0003-1436-7658]{Hansung B. Gim} %%% cosmologist.hs@gmail.com
\affiliation{Department of Physics, Montana State University, P. O. Box 173840, 
Bozeman, MT 59717, USA}

\author[0000-0001-7411-5386]{Maria del Carmen Polletta}
\affiliation{INAF--–Istituto di Astrofisica Spaziale e Fisica Cosmica Milano,  Via A.\ Corti 12, I-20133 Milano, Italy}

\author[0000-0003-3329-1337]{Seth H. Cohen} %%% seth.cohen@asu.edu
\affiliation{School of Earth and Space Exploration, Arizona State University,
Tempe, AZ 85287-1404, USA}

\author[0000-0001-9262-9997]{Christopher N.\ A.\ Willmer} %%% cnawillmer@gmail.com 
\affiliation{Steward Observatory, University of Arizona, 933 N Cherry Ave,
Tucson, AZ, 85721-0009, USA}

\author[0000-0002-7791-3671]{Xiurui Zhao}
\affiliation{Center for Astrophysics \textbar\ Harvard \& Smithsonian, 60 Garden Street, Cambridge, MA, 02138, USA}

%\author{builders here}
%\affiliation{various}

\author[0000-0002-9816-1931]{Jordan C. J. D'Silva} %%% jordan.dsilva@research.uwa.edu.au
\affiliation{International Centre for Radio Astronomy Research (ICRAR) and the
International Space Centre (ISC), The University of Western Australia, M468,
35 Stirling Highway, Crawley, WA 6009, Australia}
\affiliation{ARC Centre of Excellence for All Sky Astrophysics in 3 Dimensions
(ASTRO 3D), Australia}

\author[0000-0003-1268-5230]{Rolf A. Jansen} %%% rolfjansen.work@gmail.com
\affiliation{School of Earth and Space Exploration, Arizona State University,
Tempe, AZ 85287-1404, USA}

\author[0000-0002-6610-2048]{Anton M. Koekemoer} %%% koekemoer@stsci.edu
\affiliation{Space Telescope Science Institute,
3700 San Martin Drive, Baltimore, MD 21218, USA}

\author[0000-0002-7265-7920]{Jake Summers} %%% jakesummers7200@gmail.com
\affiliation{School of Earth and Space Exploration, Arizona State University,
Tempe, AZ 85287-1404, USA}

\author[0000-0001-8156-6281]{Rogier A. Windhorst}%%% Rogier.Windhorst@gmail.com
\affiliation{School of Earth and Space Exploration, Arizona State University,
Tempe, AZ 85287-1404, USA}

\author[0000-0001-7410-7669]{Dan Coe} %%% dcoe@stsci.edu
\affiliation{Space Telescope Science Institute, 3700 San Martin Drive, Baltimore, MD 21218, USA}
\affiliation{Association of Universities for Research in Astronomy (AURA) for the European Space Agency (ESA), STScI, Baltimore, MD 21218, USA}
\affiliation{Center for Astrophysical Sciences, Department of Physics and Astronomy, The Johns Hopkins University, 3400 N Charles St. Baltimore, MD 21218, USA}

\author[0000-0003-1949-7638]{Christopher J. Conselice} %%% conselice@gmail.com
\affiliation{Jodrell Bank Centre for Astrophysics, Alan Turing Building,
University of Manchester, Oxford Road, Manchester M13 9PL, UK}

\author[0000-0001-9491-7327]{Simon P. Driver} %%% Simon.Driver@icrar.org
\affiliation{International Centre for Radio Astronomy Research (ICRAR) and the
International Space Centre (ISC), The University of Western Australia, M468,
35 Stirling Highway, Crawley, WA 6009, Australia}

\author[0000-0003-1625-8009]{Brenda Frye} %%% brendafrye@gmail.com
\affiliation{Steward Observatory, University of Arizona, 933 N Cherry Ave,
Tucson, AZ, 85721-0009, USA}

\author[0000-0001-9440-8872]{Norman A. Grogin} %%% nagrogin@stsci.edu
\affiliation{Space Telescope Science Institute,
3700 San Martin Drive, Baltimore, MD 21218, USA}

\author[0000-0001-6434-7845]{Madeline A. Marshall} %%% madeline_marshall@outlook.com
\affiliation{National Research Council of Canada, Herzberg Astronomy \&
Astrophysics Research Centre, 5071 West Saanich Road, Victoria, BC V9E 2E7,
Canada}
\affiliation{ARC Centre of Excellence for All Sky Astrophysics in 3 Dimensions
(ASTRO 3D), Australia}

\author[0000-0001-6342-9662]{Mario Nonino} %%% nnn.mario@gmail.com
\affiliation{INAF---Osservatorio Astronomico di Trieste, Via Bazzoni 2, 34124
Trieste, Italy}

\author[0000-0002-6150-833X]{Rafael {Ortiz~III}} %%% rortizii@asu.edu
\affiliation{School of Earth and Space Exploration, Arizona State University,
Tempe, AZ 85287-1404, USA}

\author[0000-0003-3382-5941]{Nor Pirzkal} %%% npirzkal@stsci.edu
\affiliation{Space Telescope Science Institute,
3700 San Martin Drive, Baltimore, MD 21218, USA}

\author[0000-0003-0429-3579]{Aaron Robotham} %%% aaron.robotham@uwa.edu.au
\affiliation{International Centre for Radio Astronomy Research (ICRAR) and the
International Space Centre (ISC), The University of Western Australia, M468,
35 Stirling Highway, Crawley, WA 6009, Australia}

\author[0000-0001-7016-5220]{Michael J. Rutkowski} %%% michael.rutkowski@mnsu.edu
\affiliation{Minnesota State University-Mankato,  Telescope Science Institute,
TN141, Mankato MN 56001, USA}

\author[0000-0003-0894-1588]{Russell E. Ryan, Jr.} %%% rryan.asu@stsci.edu
\affiliation{Space Telescope Science Institute,
3700 San Martin Drive, Baltimore, MD 21218, USA}

\author[0000-0001-9052-9837]{Scott Tompkins} %%% satompki@asu.edu
\affiliation{School of Earth and Space Exploration, Arizona State University,
Tempe, AZ 85287-1404, USA}

\author[0000-0001-7592-7714]{Haojing Yan} %%% yanhaojing@gmail.com
\affiliation{Department of Physics and Astronomy, University of Missouri,
Columbia, MO 65211, USA}

\author[0000-0001-8751-3463]{Heidi B.~Hammel} % hbhammel@aura-astronomy.org
\affiliation{Association of Universities for Research in Astronomy, 1331 Pennsylvania Avenue NW, Suite 1475, Washington, DC 20005, USA}

\author[0000-0001-7694-4129]{Stefanie N.~Milam} % stefanie.n.milam@nasa.gov
\affiliation{NASA Goddard Space Flight Center, Greenbelt, MD\,20771, USA}

\author[0000-0003-4875-6272]{Nathan J.\  Adams}
\affiliation{Jodrell Bank Centre for Astrophysics, Alan Turing Building,
University of Manchester, Oxford Road, Manchester M13 9PL, UK}

\author[0000-0002-0005-2631]{John F. Beacom}  %%% beacom.7@osu.edu
\affiliation{Center for Cosmology and AstroParticle Physics, 191 W. Woodruff Avenue, Columbus, OH 43210, USA}
\affiliation{Department of Physics, Ohio State University, 191 W. Woodruff Avenue, Columbus, OH 43210, USA}
\affiliation{Department of Astronomy, Ohio State University, 140 West 18$^{th}$ Avenue, Columbus, OH 43210, USA}

\author[0000-0003-0883-2226]{Rachana Bhatawdekar}
\affiliation{European Space Agency (ESA), European Space Astronomy Centre (ESAC), Camino Bajo del Castillo s/n, 28692 Villanueva de la Cañada, Madrid, Spain}

\author[0000-0003-0202-0534]{Cheng Cheng}
\affiliation{Chinese Academy of Sciences South America Center for Astronomy, National Astronomical Observatories, CAS, Beijing, 100101, China}

\author{F.\ Civano}
\affiliation{Center for Astrophysics \textbar\ Harvard \& Smithsonian, 60 Garden Street, Cambridge, MA, 02138, USA}

\author[0000-0001-7363-6489]{W.\ Cotton}
\affiliation{National Radio Astronomy Observatory (NRAO), 520 Edgemont Road, Charlottesville, VA 22903, USA}

\author[0000-0003-4738-4251]{Minhee Hyun}
\affiliation{Korea Astronomy and Space Science Institute, 776 Daedeok-daero, Yuseong-gu, Daejeon 34055, Republic of Korea}

\author[0000-0003-3214-9128]{Satoshi Kikuta}
\affiliation{National Astronomical Observatory of Japan, 2-21-1,
  Osawa, Mitaka, Tokyo 181-8588 Japan}

\author{K. E. Nyland}
\affiliation{U.S.\ Naval Research Laboratory, 4555 Overlook Ave SW, Washington, DC 20375, USA}

\author[0000-0002-5187-7107]{W. M. Peters}
\affiliation{U.S.\ Naval Research Laboratory, 4555 Overlook Ave SW, Washington, DC 20375, USA}

\author[0000-0003-4030-3455]{Andreea Petric} %%% apetric@stsci.edu
\affiliation{Space Telescope Science Institute, 3700 San Martin Drive,
Baltimore, MD 21218, USA}

\author[0000-0001-8887-2257]{Huub J. A. R\"ottgering} %%% huubrottgering@gmail.com 
\affiliation{Leiden Observatory, PO Box 9513, 2300 RA Leiden, The Netherlands}

\author[0000-0001-5648-9069]{T.\ Shimwell}
\affiliation{ASTRON, the Netherlands Institute for Radio Astronomy, Postbus 2, NL-7990 AA, Dwingeloo, the Netherlands}
\affiliation{Leiden Observatory, Leiden University, PO Box 9513, NL-2300 RA Leiden, the Netherlands}

\author[0000-0001-7095-7543]{Min S.\ Yun} %%% myunm82@gmail.com 
\affiliation{Department of Astronomy, University of Massachusetts, Amherst, MA
01003, USA}


\begin{abstract}
The Time Domain Field (TDF) near the North Ecliptic Pole in JWST's continuous-viewing zone will become a premier ``blank field'' for extragalactic science.  JWST/NIRCam data in a 16\,arcmin$^2$ portion of the TDF identify 4.4\,\micron\ counterparts for 62 of 63 3\,GHz sources with $S(\rm 3\,GHz)>5$\,\mmJy.  The one unidentified radio source may be a lobe of a nearby Seyfert galaxy, or it may be an infrared-faint radio source.  The bulk properties of the radio-host galaxies are consistent with those found by previous work: redshifts range from 0.14 to 4.4 with a median redshift of 1.33. The radio emission arises primarily from star formation in $\sim$2/3 of the sample and from an active galactic nucleus in $\sim$1/3, but just over half the sample shows evidence for an AGN either in the spectral energy distribution or by radio excess.  All but three counterparts are brighter than magnitude 23~AB at 4.4~\micron, and the exquisite resolution of JWST identifies correct counterparts for sources for which observations with lower angular resolution would mis-identify a nearby bright source as the counterpart when the correct one is faint and red.  \edit1{Up to 11\% of counterparts might have been unidentified or misidentified absent NIRCam observations.}
\end{abstract}

\section{Introduction} \label{s:intro}

The history of energy generation in the Universe includes both accretion, especially onto supermassive black holes, and star formation.  Early radio surveys \edit1{with the NSF's Karl G.\ Jansky Very Large Array (VLA)\footnote{The National Radio Astronomy Observatory is a facility of the National Science Foundation operated under cooperative agreement by Associated Universities, Inc.}} \citep[\eg, FIRST, NVSS;][]{Becker94,Condon98} explored mainly luminous radio galaxies, powered by accretion onto supermassive black holes, because such sources could be detected at enormous distances 
\citep[\eg,][]{Chambers1996,Seymour2007,Miley2008,Saxena2019}. When radio surveys began to reach flux densities near 1\,mJy, a significant fraction of radio sources turned out to be starburst galaxies \citep[\eg,][]{Windhorst1985}.  
Modern, far deeper, radio surveys probe both types of energy generation and complement optical studies of cosmic star formation and accretion
\citep[\eg,][and references therein]{Smolcic2017data,Algera2020,Tompkins2023}. 
Broad multi-wavelength (and multi-messenger) coverage leads to better understanding of sources'  underlying physical processes.

A limitation of radio continuum surveys is that they do not provide redshifts and therefore distances.  For distances, it is necessary to identify visible or infrared counterparts of the radio sources and obtain redshifts, ideally spectroscopic but at least photometric.  Finding counterparts is not always as easy as a simple positional match.  Radio source positions have historically been uncertain because of large beam sizes, and for powerful double-lobe radio sources, the host galaxy is between the lobes rather than coincident with either of them.  \citet{Spinrad1985} provided nearly complete identifications for the extragalactic sources in the 3CR, a survey complete to 8~Jy at 178~MHz. At fainter radio flux densities, the optical identification fraction was lower, about $\sim$75--85\%  in early \edit1{Hubble Space Telescope} (HST) imaging surveys \citep[\eg,][]{Fomalont2006, Russell2008, Windhorst1995}. More recent, deeper surveys  have identification completeness above 90\% \citep[\eg][]{Smolcic2017id,Owen2018,Algera2020,Kondapally2021}, even as radio flux densities have reached \mmJy\ levels.

To reach very high completeness levels, observations \edit1{at wavelengths $\ga$2\,\micron} are essential because passive stellar populations are intrinsically red, and star-forming galaxies can be reddened by dust.  High redshift increases the need for infrared observations because light observed at visible wavelengths originated as rest-frame ultraviolet (UV) emission, which is often absent or severely reddened.  The stellar emission from galaxies typically peaks near 1.6~\micron\ \citep{Sawicki02}, and observing longward of that means the increasing intrinsic flux density towards shorter wavelengths somewhat cancels the effect of redshift (``negative K correction''). Another benefit of longer wavelengths is that dust extinction is much smaller. 
As an example, \citet{Strazzullo2010} found optical ($gri$-bands) counterparts for $\ge$83\% of a sample of $S(\rm 1.4\,GHz) > 13.5$\,\mmJy\ (5$\sigma$) radio sources. Adding \textit{JHK} data raised the identification rate to $\ge$90\%, and adding IRAC 3.6 and 4.5\,\micron\ data (reaching mags $\sim$24.3) raised it to $\ge$93\%. Sources added at the longer wavelengths were mostly at $z>1.3$.  \citet{Smolcic2017id} achieved a 90\% identification rate for the VLA-COSMOS 3\,GHz radio survey, which reached rms sensitivity of 2.3~\mmJy.  This was in the COSMOS field, which has deep HST data and additional data in many wavebands.  Adding IRAC 3.6~\micron\ data raised the completeness rate to 92.4\%.  More dramatically, \citet{Willner2012} used 4.5~\micron\ IRAC observations to identify 100\% of radio sources in  1.4~GHz survey \citep{Ivison2007} in the Extended Groth Strip, though the radio survey depth was only 50~\mmJy.  More recently, \citet{Cotton2018} found 4.5\,\micron\ counterparts for 98\% of 3\,GHz sources reaching 4\,\mmJy.

With the advent of JWST, it seems natural to investigate how well it can identify counterparts of faint radio sources and what JWST's exquisite angular resolution can reveal about their morphologies.  This paper reports an initial attempt in the Prime Extragalactic Areas for Reionization and Lensing Science (PEARLS) Time Domain Field (TDF). This field is near the North Ecliptic Pole and within JWST's continuous viewing zone, making observations possible at any time. 
The observational design and layout of the original JWST TDF were described by \citet{Jansen2018}.  Its specific location was selected to minimize the number of bright stars in the field.
\citet{Windhorst2023} gave full details of the JWST observations (PID 2738, PIs R.\ Windhorst and H.\ Hammel) and data.\footnote{The latest news about the field is at\\ \url{http://lambda.la.asu.edu/jwst/neptdf/}.}
Briefly, the TDF area consists of four ``spokes,'' each subtending an intended area $2\farcm3\times 7\farcm4$ at orientation differences of 90\arcdeg. Figure~1 of \citet{Willmer2023} depicts the field layout and location.
The radio observations \citep{Hyun2023} were centered on a quasar located about 1\arcmin\ south of the bright star at the intersection of the southern and eastern spokes.

This paper uses the JWST/NIRCam images of the first spoke observed in the TDF to identify counterparts of 3\,GHz radio sources from the \citet{Hyun2023} sample. The main goal is to describe the identification process and results. The paper is organized as follows: Section~\ref{s:obs} describes the radio survey, the JWST observations used for the identifications, and additional data used.  Section~\ref{s:match} describes the matching procedure and results.  Section~\ref{s:disc} gives the source and sample characteristics, and Section~\ref{s:summ} summarizes the results, including recommendations for future matching studies.
Distances are based on flat $\Lambda$CDM cosmology with $H_0=70.4$\,km\,s$^{-1}$\,Mpc$^{-1}$ and $\Omega _{\rm M} =0.272$ \citep{Komatsu2011}.  All magnitudes are AB.

\section{Observations} \label{s:obs}
\subsection{Radio observations}
\label{s:radio}
\citet[][their Appendix~A]{Hyun2023} described the 3~GHz radio observations used here. The observations used the A and B arrays \edit1{(44 and 4 hours, respectively)} of the VLA to cover a single field centered on the $z=1.4409$ quasar ICRF J172314.1+654746.  The field is 12\arcmin\ in radius, at which distance the VLA primary-beam response falls to 0.1.  The angular resolution is 0\farcs7, and the rms noise near the beam center is 1\,$\mu$Jy\,beam$^{-1}$. \edit1{Away from the beam center, the noise grows as the reciprocal of primary-beam response.}  \citet{Hyun2023} extracted 756 $\ge$5$\sigma$ sources from the image, and that list (their Table~3) is the input source list for this paper.  
\edit1{Because of the JWST coverage (Section~\ref{s:jwst}), all sources studied here are} within 6\farcm25 of the beam center, where the primary-beam response is \edit1{$\ge$0.61}.
The radio spectral bandwidth was 2.0\,GHz, but lack of precise knowledge of the frequency-dependent primary-beam correction  prevents derivation of meaningful in-band spectral indices.
Compared to other recent 3\,GHz surveys, the observations are more sensitive than the VLA-COSMOS \citep{Smolcic2017data} survey, comparable to the Lockman Hole \citep{Cotton2018} and the VLA Frontier \citep{Heywood2021} surveys, but less sensitive than COSMOS-XS \citep{Algera2020}.
 
The radio sources studied here are small in size because of the VLA observations used to select them. \edit1{The high angular resolution means that extended, low-surface-brightness sources may  be missed.} A survey in the Lockman Hole field with similar sensitivity but including VLA C-array data found a median source size 0\farcs3 \citep{Cotton2018}. About 2/3 of the \citet{Hyun2023} sources are unresolved, consistent with the \citeauthor{Cotton2018} median size estimate. 
However, the \citeauthor{Cotton2018} results suggest that adding C-array data would have found about 10\% more sources than the \citeauthor{Hyun2023} A+B-array data did.  These are presumably extended galaxies with low surface brightness 
\edit1{that are not included  in the existing radio catalog. This paper's objective is to examine the radio-source population detected by typical high-resolution radio surveys.  However, if the missing radio sources are low-redshift spiral or irregular galaxies, JWST should have no trouble detecting them. Appendix~\ref{sa:sb} provides some estimates of the number of such sources.}
Another issue is that large sources may have their flux densities underestimated.
The largest source studied here has a deconvolved \citep{Hyun2023} major-axis diameter of 2\farcs0, so this is probably not a major problem in the present study.

\subsection{JWST observations}
\label{s:jwst}

The first spoke of the TDF was observed with JWST/NIRCam in 2022 July. Each spoke comprises two separate observation blocks, offset from one another spatially.  Because of a guide star issue, the two blocks were observed 12 days apart at an orientation difference of about 8\arcdeg, leading to a ``dogleg'' in the area covered, which is about $2\farcm3\times7\farcm0$.  The three remaining spokes have now been observed, but those data are not public and are not used in this paper.  Parallel NIRISS observations were also obtained, but the public NIRISS data taken at the time of Spoke~1 are in the Spoke~3 NIRCam area and are not discussed here. The NIRCam filters observed were F090W, F115W, F150W, F200W, F277W, F356W, F410M, and F444W.   Image quality is diffraction-limited \citep{Rigby2023} with FWHM ranging from 60 to 160 milliarcseconds (mas) at the wavelengths observed. \edit1{Nominal exposure times at each sky location were 2920\,s for four filters and 3350\,s for the other four.}  Point-source sensitivities range from 28.1 to 29.1\,mag (5$\sigma$); per-filter values and
further details are given by \citet[][\edit1{their Table~2}]{Windhorst2023}.

All images were calibrated using version 1.7.2 of the JWST pipeline\footnote{\url{https://github.com/spacetelescope/jwst}} \edit1{\citep{pipeline}} with reference files specified by the context file jwst\_0995.pmap.  Final mosaics in all filters have pixel scales of 30 
mas/pixel.  All the mosaics were aligned to the GAIA-DR3 \citep{gaia3}\footnote{\url{https://www.cosmos.esa.int/web/gaia/}} astrometric reference frame. The four short-wavelength (SW) filters covered identical areas, as did the four long-wavelength (LW) filters, but the SW and LW areas are offset by $\sim$2\arcsec. 

\subsection{Ground-based imaging}
\label{s:ground}

The entire TDF was imaged with the Subaru Hyper-Suprime-Cam (HSC) as
part of the Hawaii EROsita Ecliptic pole Survey (HEROES; e.g., \citealt{Songaila2018,Taylor2023}). Those images were publicly available in early 2020 on the SMOKA archive of the National Astronomical Observatories of Japan.
\citet{Willmer2023} described the source extraction and photometry. \citet{Willmer2023} also reported observations with the MMT and Magellan Infrared Spectrograph (MMIRS;  \citealt{McLeod2012,Chilingarian2015}) on the MMT.
The data used here comprise $g$, $i2$, and $z$ from HSC and $YJHK$ from MMIRS.

\subsection{Spectroscopic redshifts}
The principal spectroscopy came from BINOSPEC
\citep{Fabricant2019} at the MMT Observatory. Willmer \etal\ (in preparation) will give full details, but in brief, the observations used a
270\,lines/mm grating to achieve spectral coverage from 
4000\,\AA\ to  9000\,\AA\ with a typical dispersion of 1.30\,\AA/pixel and
resolution of 1340. The sample
of 209 galaxies observed, limited to $r\la23$, was selected from two catalogs: a preliminary catalog coming from the HEROES Subaru/HSC imaging (Hasinger, private communication) and
a catalog derived from MMT/MMIRS \textit{YJHK} imaging \citep{Willmer2023}.  \citet{Hyun2023} have already published some of the measurements, specifically those for SCUBA2 sources.

The field was also observed by Hectospec \citep{Fabricant2005} at the MMT on 2023 May 21 (PIs F.~Civano and X.~Zhao).
Hectospec is fiber-fed with each fiber subtending 1\farcs5. The observations used a 270\,lines/mm grating to achieve spectral coverage from 3650 to 9200\,\AA\ with typical dispersion of 1.2\,\AA/pixel and spectral resolution 6\,\AA.  The galaxies targeted were primarily XMM X-ray sources.
These observations confirmed \edit1{five of the Binospec redshifts including improving $Q=3$ to $Q=4$ for two sources.}

\section{Source matching} 
\label{s:match}
\subsection{Match process}
\label{s:process}

\begin{figure*}
{\bf ~~~~3\,GHz\hfill ~~F090W \hfill F115W \hfill F150W \hfill F200W \hfill F277W \hfill F356W \hfill F400M~ \hfill F444W~~~~}\\
\includegraphics[width=\linewidth]{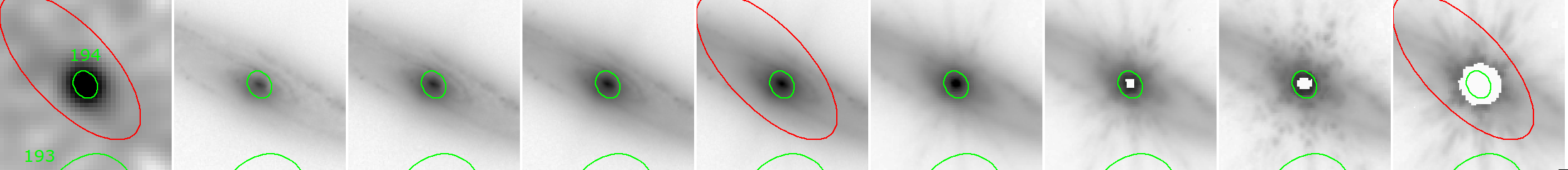}\\
\includegraphics[width=\linewidth]{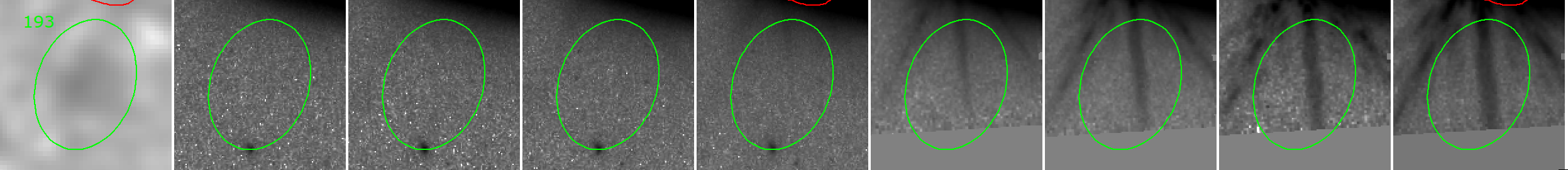}\\
\includegraphics[width=\linewidth]{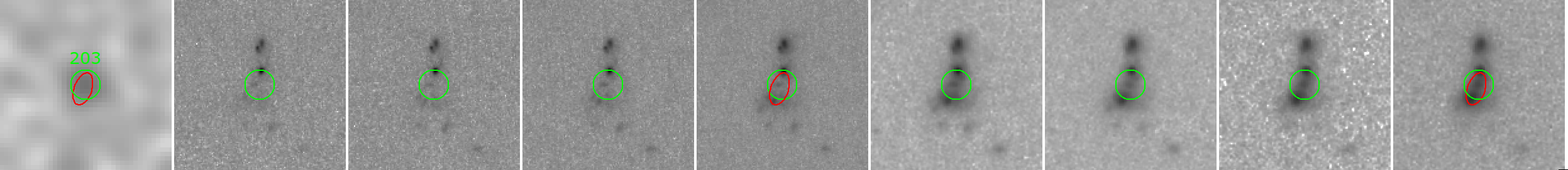}\\
\includegraphics[width=\linewidth]{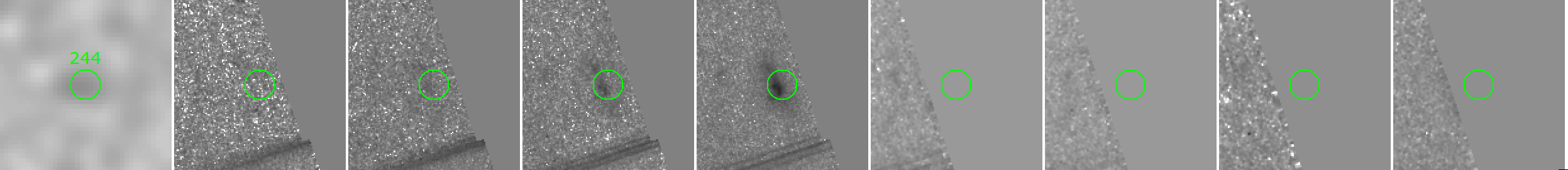}\\
\includegraphics[width=\linewidth]{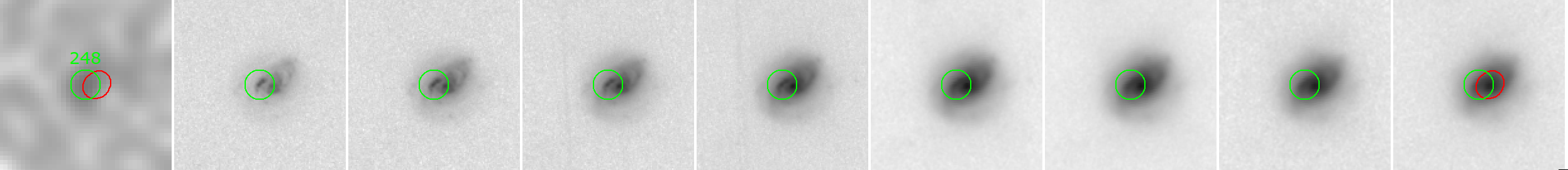}\\
\includegraphics[width=\linewidth]{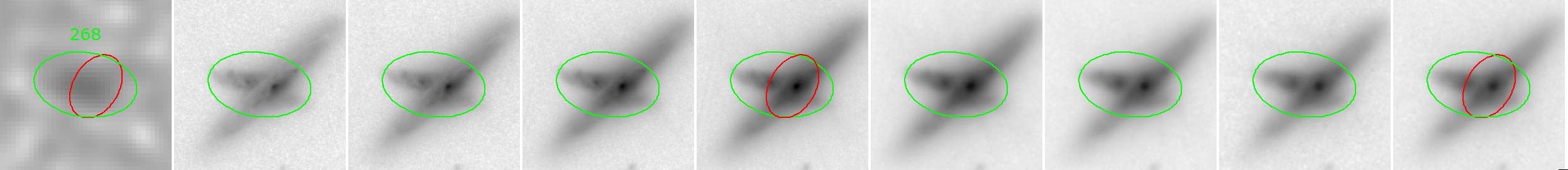}\\
\includegraphics[width=\linewidth]{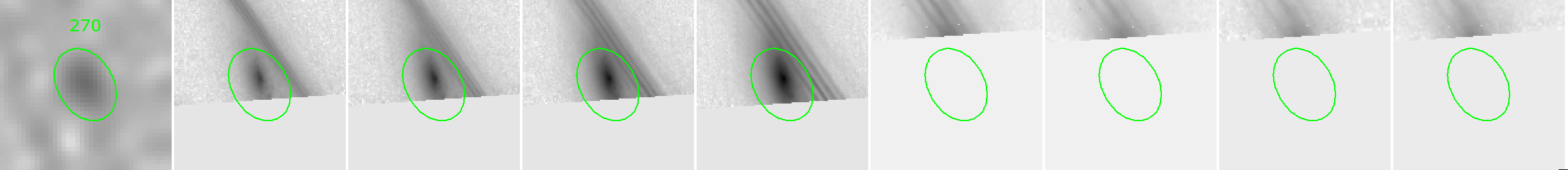}\\
\includegraphics[width=\linewidth]{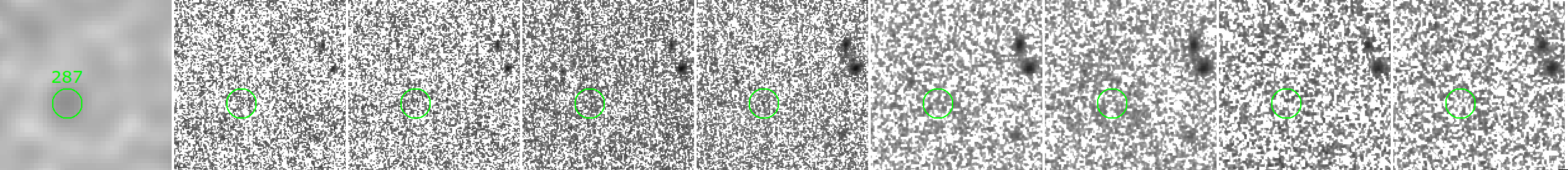}\\
\includegraphics[width=\linewidth]{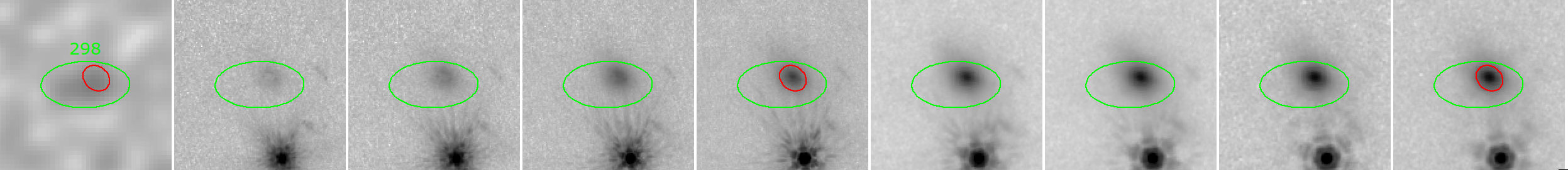}\\
\includegraphics[width=\linewidth]{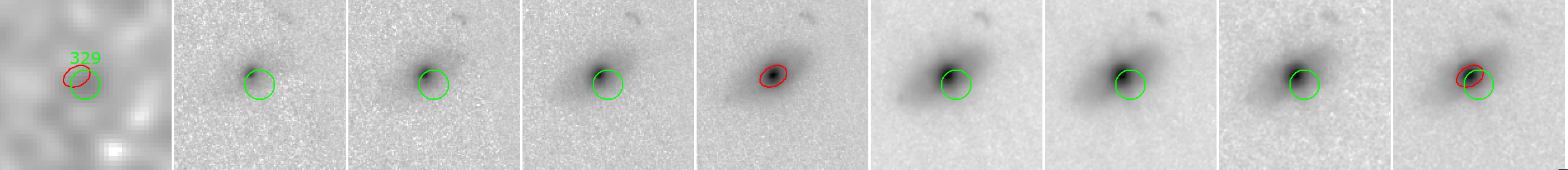}
\caption{Negative images of ten sources for which the automated matching failed to find a counterpart within 0\farcs24.  Leftmost panels show the 3\,GHz radio image with the source ID included. Other panels show the NIRCam images in order of wavelength as labeled at the top.  Each thumbnail is 4\arcsec\ on a side with north up, east to the left. Radio images are all on the same greyscale ($-3$ to +35\,\mmJy\,beam$^{-1}$) and are before primary-beam correction to give nearly constant noise.  NIR images have different greyscales for each source, but all images for a given source have the same greyscale to show color information.
Green circles or ellipses show the radio positions and radio source sizes \citep{Hyun2023}.  Red ellipses, plotted on only three images of each source to avoid hiding details, show F444W source parameters from SExtractor. (MAG\_AUTO photometry ellipses are about 2.5 times larger than the ellipses shown.) The ID~193 images show the outskirts of ID~194 on the north edge and, in the long-wave images, diffraction spikes of the ID~194 nucleus.}
\label{f:bigsep}
\end{figure*}

\begin{figure*}
{\bf ~~~~3\,GHz\hfill ~~F090W \hfill F115W \hfill F150W \hfill F200W \hfill F277W \hfill F356W \hfill F400M~ \hfill F444W~~~~}\\
\includegraphics[width=\linewidth]{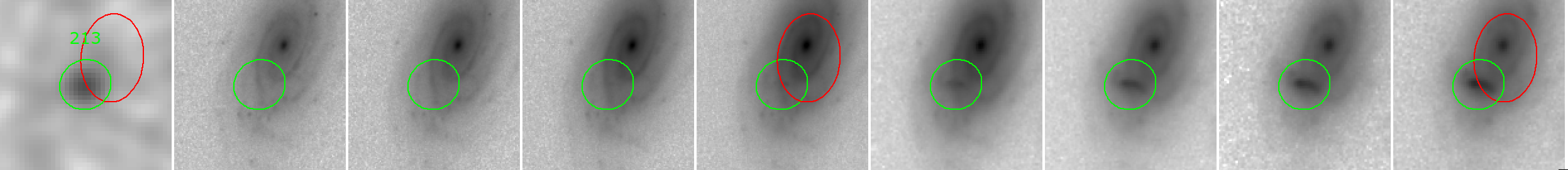}\\
\includegraphics[width=\linewidth]{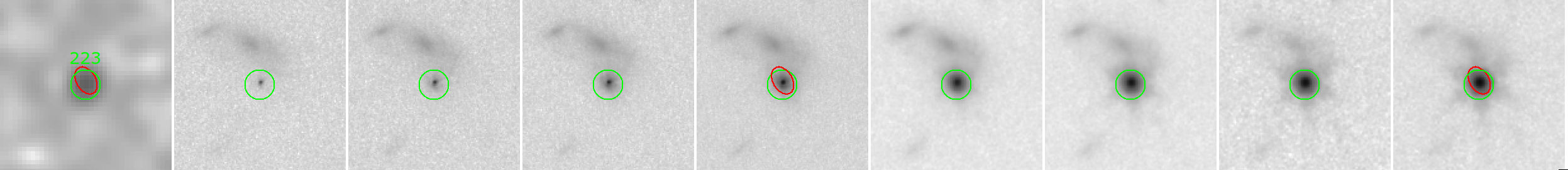}\\
\includegraphics[width=\linewidth]{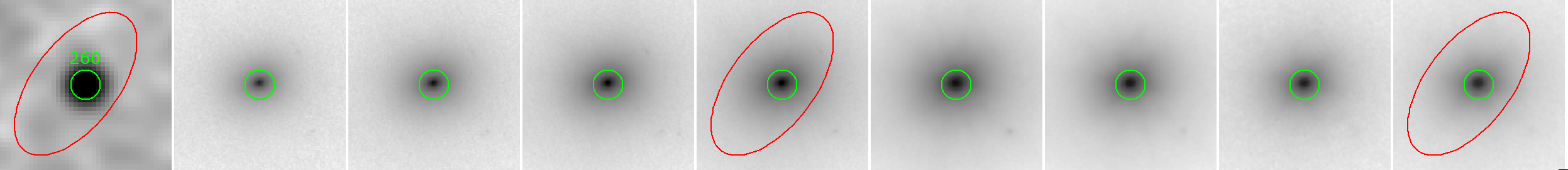}\\
\includegraphics[width=\linewidth]{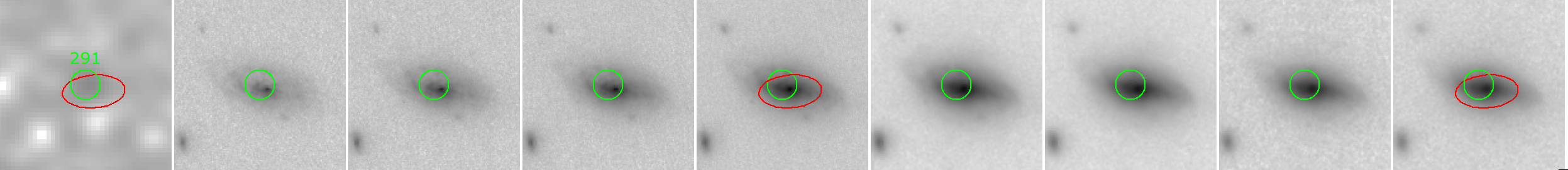}\\
\includegraphics[width=\linewidth]{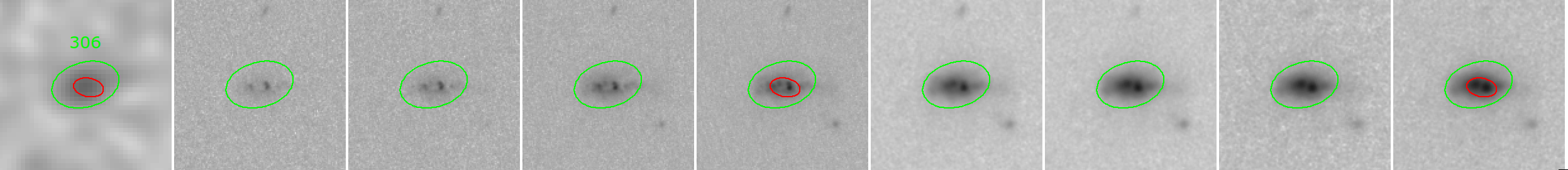}\\
\includegraphics[width=\linewidth]{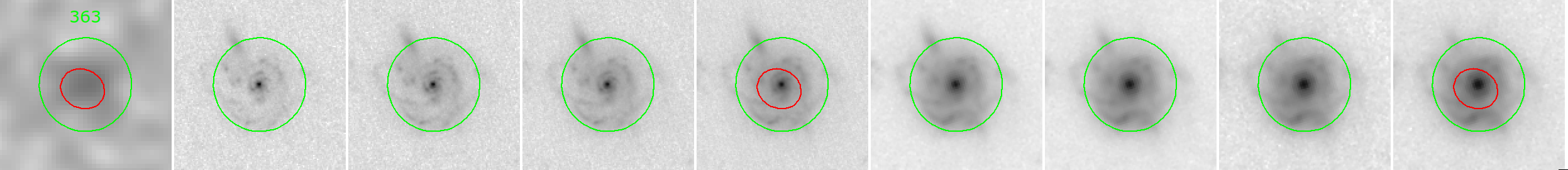}\\
\caption{Negative images of six interesting sources.  Leftmost panels show the 3\,GHz radio image with the source ID included. Other panels show the NIRCam images in order of wavelength as labeled at the top.  Each thumbnail is 4\arcsec\ on a side with north up, east to the left. Radio images are all on the same greyscale ($-3$ to +35\,\mmJy\,beam$^{-1}$) and are before primary-beam correction to give nearly constant noise.  NIR images have different greyscales for each source, but all images for a given source have the same greyscale to show color information.
Green circles or ellipses show the radio positions and radio source sizes \citep{Hyun2023}.  Red ellipses show F444W source parameters from SExtractor. (MAG\_AUTO photometry ellipses are about 2.5 times larger than the ellipses shown.) }
\label{f:tricky}
\end{figure*}

An initial NIRCam object catalog was created with SExtractor \citep{Bertin1996} on the F444W image.  Then for each radio source, the area around it and around the nearest 4.4~\micron\ source was examined visually in all images to verify NIRCam coverage and look for  counterparts. 
One source, ID~255 in the \citet{Hyun2023} list, is just off the edge of the SW coverage and well outside the LW coverage.  Its counterpart is apparently an extended galaxy because a partial image with hints of a spiral arm can be seen in the SW pixels nearest the edge, but this source is excluded from the present sample.
Two radio sources (ID~244/270) are in the area that has SW but not LW observations.
These are included in the sample, but their photometry in the SW filters is less reliable than for most sources because part of their light is outside the image.  The initial search thus gave a sample of 64 radio sources.

\begin{figure*}[tb!]
\includegraphics[width=\linewidth]
{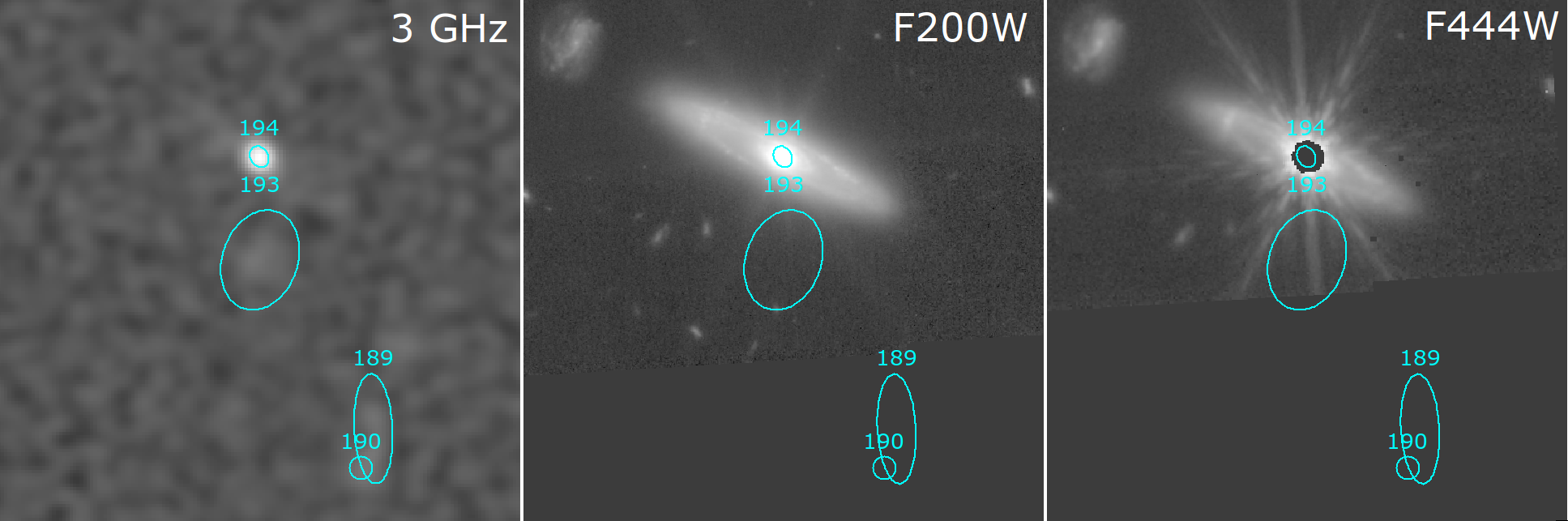}
\caption{Wider-field images of ID~193. The radio image is on the left, and the other panels show two NIRCam images as labeled. Field of view is 16\arcsec\ in all panels. Ellipses show positions and sizes of four radio sources, the southern two of which are outside the NIRCam coverage.  The ID~194 nucleus is saturated in the original NIRCam data and therefore appears black.}
\label{f:193}
\end{figure*}

For 62 of the 64 radio sources in the NIRCam field, there is an obvious infrared source within 0\farcs5. One of the exceptions is ID~193, discussed in Section~\ref{s:missing}.  The other is ID~287, the faintest radio source among the 64. ($S(\rm 3\,GHz)=4.7\pm0.9$\,\mmJy\ with the next-faintest at $5.4\pm1.1$\,\mmJy.)  As shown in Figure~\ref{f:bigsep}, no counterpart is visible in any NIRCam image.  The 5$\sigma$ upper limit is 28.35\,mag \citep{Windhorst2023}, and a source 2\farcs4 to the northwest, 27.3\,mag in F444W, is easily visible at all wavelengths. \citet{Hyun2023} cleaned their image only down to 5\,\mmJy\ \edit1{(5.3\,\mmJy\ at this position after primary-beam correction)}, and indeed the radio image of the source (Figure~\ref{f:bigsep}) is less than compelling.  In order to check the reality of ID~287, we performed an independent  search on the same 3\,GHz image as \citeauthor{Hyun2023} used. Source detection used Python Blob Detection and Source Finder \citep[PyBDSF;][]{Mohan2015} with a detection threshold 5$\sigma$ above the local rms noise  calculated by the program. That search found a few sources that \citeauthor{Hyun2023}\ did not, but overall it found only 708 sources on the image compared to 756 by \citeauthor{Hyun2023}.  ID~287 was one of the ones not found.  For the rest of this paper, we will consider only the 63 sources with $S(\rm 3\,GHz)>5$\,\mmJy. If ID~287 is a real source, our completeness numbers should be multiplied by 63/64.

\begin{figure}
\includegraphics[width=\linewidth, clip=true, trim=20 78 40 0]{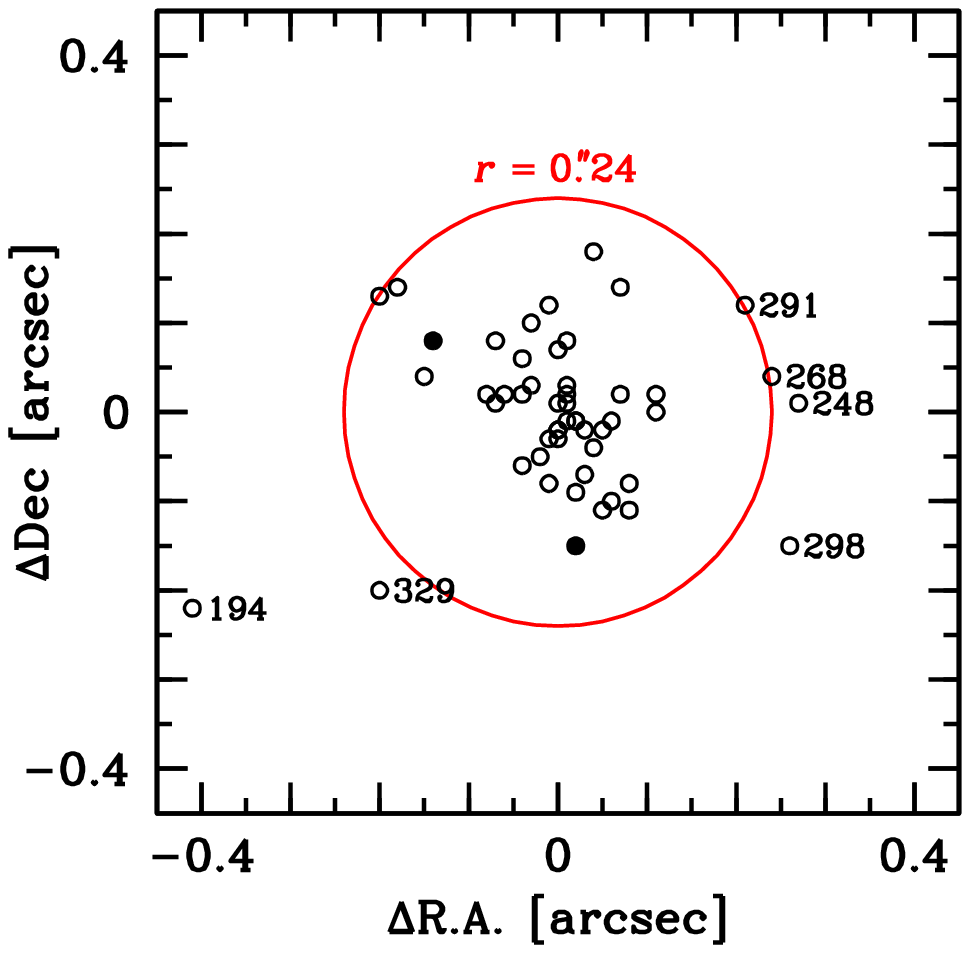}
\caption{Offsets between radio catalog position and F444W position from SExtractor.  Filled circles indicate the two sources outside the F444W image and measured on the F200W image. The red circle has radius 0\farcs24, and the radio IDs are shown for objects with larger offsets than this.  The nucleus of ID~194, which is saturated on the F444W image, can be measured on the F200W image and is only 0\farcs03 from the radio position.}
\label{f:xy}
\end{figure}

Automated \edit1{(nearest-neighbor)} matching found an F444W source within 0\farcs24 of the radio position in 54 cases. 
For all of these, visual inspection showed an obvious identification, usually a galaxy well resolved on the NIRCam image. The only incorrect identification is ID~213 (Figure~\ref{f:tricky}), which is near, or perhaps within, a large spiral galaxy.  However, the radio counterpart's red colors suggest that the counterpart is actually a background galaxy seen through the spiral's disk. ID~306 shows multiple clumps, the reddest of which coincides with the radio position.  Searches at wavelengths $<$2.0~\micron\ would not have identified the correct counterpart for either ID~213 or ID~306.  However, a position search at 4.4\,\micron\ with a match radius of 0\farcs24 would have correctly identified 54 of 61 or 89\% of possible counterparts in this sample.

The two objects outside the LW image (ID~244/270) are visible on the SW images. (The automated F444W search found unrelated sources 0\farcs7 and 1\farcs8 away, respectively.)  A similar search on the F200W image would have found the correct counterpart for ID~244 with separation 0\farcs16.  The counterpart is faint and red and might not have been seen at wavelengths $<$2~\micron. ID~270, in addition to being near the SW image edge, is near some diffraction spikes from a bright star outside the image, and  SExtractor found the spikes rather than the counterpart galaxy.  The derived ``source'' (diffraction spikes) major-axis length is 7\farcs5, so the offset from the radio source could have been large, but by luck it was only 0\farcs42.  The actual counterpart to ID~270 is bright and extended and separated only 0\farcs18 from the radio position. Identifications look reliable for both sources.  In particular, there is no reason to believe an F444W image would show different counterparts.

The remaining six sources with radio to F444W offsets $>$0\farcs24 are a mixed group as shown in Figure~\ref{f:bigsep}.  They are discussed individually in Section~\ref{s:special}, but all the final counterpart identifications look reliable as discussed further in Section~\ref{s:reliability}.

\subsection{Missing source}
\label{s:missing}

One radio source, ID~193, has no visible counterpart in the NIRCam images.  Its 3\,GHz flux density is 21\,\mmJy\ \edit1{(13\,\mmJy\ prior to primary-beam correction)} with 4.9$\sigma$ detection significance \citep{Hyun2023} \edit1{or 5.5$\sigma$ by PyBDSF.} The radio source is extended but of uncertain size with a major axis diameter 1\farcs2$\pm$0\farcs3 and only an upper limit (1\farcs6) for the minor axis diameter. \edit1{The uncertainty in source size contributes to the flux-density uncertainty but does not diminish the detection significance.} ID~193 is 3\farcs2 south of the bright Seyfert galaxy ID~194, but 194's 3\,GHz flux density is only 81\,\mmJy, and the radio image (Figure~\ref{f:193}) shows no evidence of sidelobes from 194 or any other bright source.  There is no reason to doubt the radio detection.

\begin{figure}
\includegraphics[width=\linewidth, clip=true, trim=36 268 80 30]
{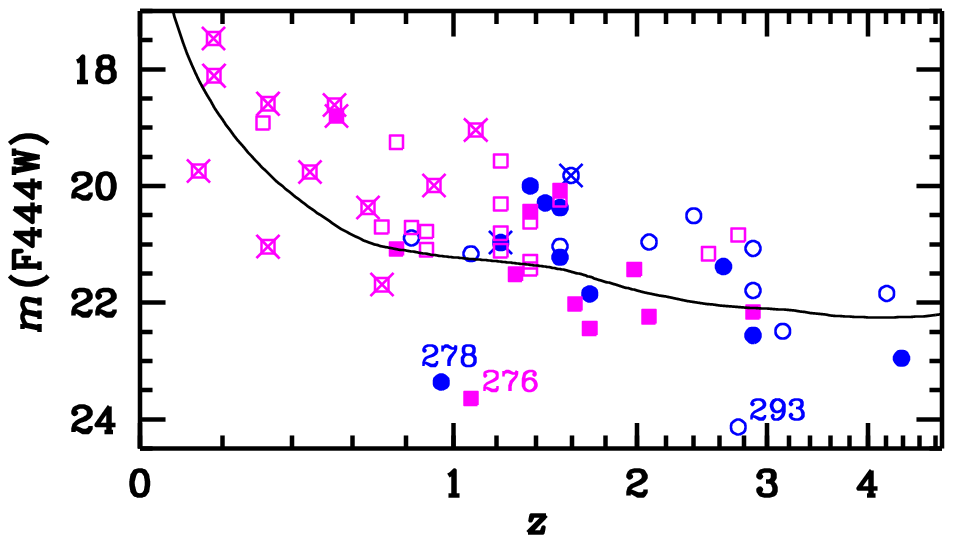}
\caption{F444W apparent AB magnitude versus redshift scaled as $\log(1+z)$ for the host galaxies. 
Point shapes and colors indicate SED classes defined in Section~\ref{s:nature}: blue circles represent type QSO, and magenta squares represent type Gal.  Filled symbols indicate galaxies with excess radio emission as defined in Section~\ref{s:properties}, and $\times$ symbols indicate galaxies with spectroscopic redshifts. 
Three faint sources are labeled.
The solid line shows the F444W magnitude of a stellar population (\citealt{Bruzual2003} stellar models) with stellar mass $10^{11}$\,\Msol\ that formed (with ${\rm SFR}\propto \exp[-t/0.5\,{\rm Gyr]}$) at $z=7$.
}
\label{f:fz}
\end{figure}

As Figure~\ref{f:193} shows, the sensitivity of the F444W image is compromised by diffraction spikes from the saturated nucleus of ID~194.  The highest surface brightness near ID~193's position is about  2\,nJy\,pixel$^{-1}$, and a source at this surface brightness occupying 10 pixels ought to be seen.  The upper limit is therefore 28\,mag\,AB, only modestly worse than the nominal sensitivity limit of 28.35 \citep{Windhorst2023}.  The faintest radio counterpart detected is ID~293 at 24.13\,mag, a factor of 35 brighter than the detection limit at ID~193's position. Even better upper limits apply in other  NIRCam filters.  The non-detections give an upper limit on the stellar mass of a host galaxy of ${\la}1.5\times10^8$\,\Msol\ at any $z<5$ (Figure~\ref{f:fz}) unless the stellar population is heavily reddened at rest NIR wavelengths.

One strong possibility for the missing ID~193 counterpart 
is that the 3\,GHz source is a radio lobe of the Seyfert galaxy ID~194. At ID~194's distance, the offset corresponds to 9.6\,kpc projected distance from the nucleus, comparable to the radius of the galaxy's disk, 6.6\,kpc. Most radio lobes have steep radio spectra \citep{Blundell2000}, and sensitive low-frequency radio maps sometimes show a jet connecting the nucleus to the lobe. 
The VLA Low-band Ionosphere and Transient Experiment (VLITE; \citealt{vlite}) is a commensal instrument on the VLA. It observed the TDF at 340\,MHz simultaneously with the \citet{Hyun2023} 3\,GHz observations.
The VLITE data were reduced independently for this paper. 
At ID~193's position, the rms noise is 140~$\mu$Jy\,beam$^{-1}$.
The VLITE angular resolution is 5\farcs6$\times$3\farcs7 with the major axis oriented east--west, thereby marginally separating IDs~193 and~194.  ID~194 was detected at 630~\mmJy\ ($\sim$4.5$\sigma$), giving a spectral index $\alpha\approx-0.9$.\footnote{$S_\nu\propto\nu^\alpha$}  The VLITE image shows  no  sign of ID~193, which at  3$\sigma$ requires spectral index flatter than $-1.4$.  Another low-frequency observation is from LOFAR \citep{lofar,Shimwell2022}.  The LOFAR 144\,MHz image has a beam size of 6\arcsec\ and therefore cannot reliably separate ID~193 from~194 or even from ID~189/190, which are $\sim$10\arcsec\ southwest of 193/194 but outside the NIRCam image. However, the LOFAR emission is slightly extended and is best fit with two Gaussians, one centered on ID~194 (but overlapping ID~193) and the other centered on ID~189/190. Fitting gives a total flux density near ID~194 of 1390\,\mmJy, implying a very rough estimate for that source of $\alpha\sim -0.9$ between 144 and 340\,MHz.  Though a contribution from ID~193 is not ruled out, the data do not require any flux from that source.  The radio results thus show no evidence for a steep-spectrum component, but the upper limits neither confirm nor rule out the possibility that ID~193 is a radio lobe.

\edit1{
\begin{figure}
\includegraphics[width=0.495\linewidth]{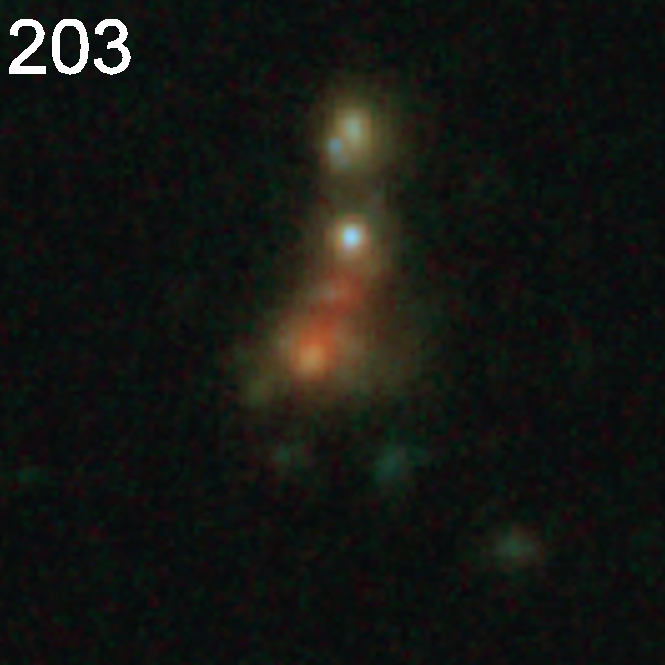}
\hfill
\includegraphics[width=0.495\linewidth]{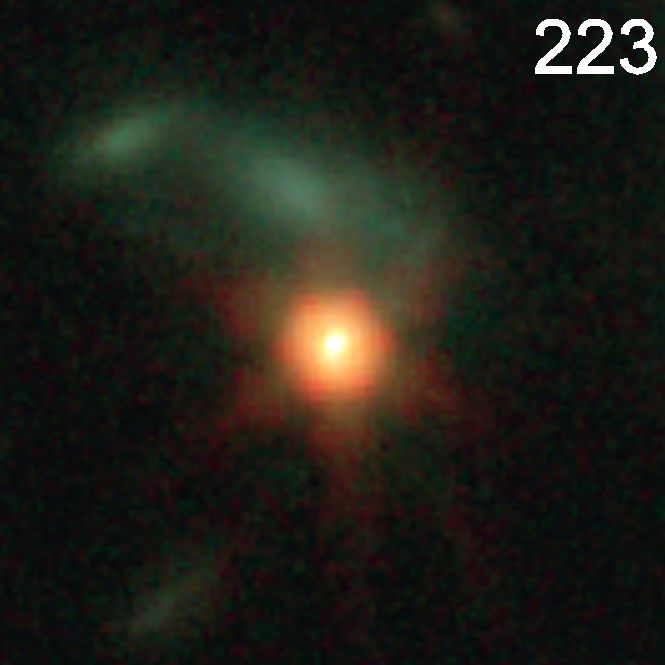}\\
\includegraphics[width=0.495\linewidth]{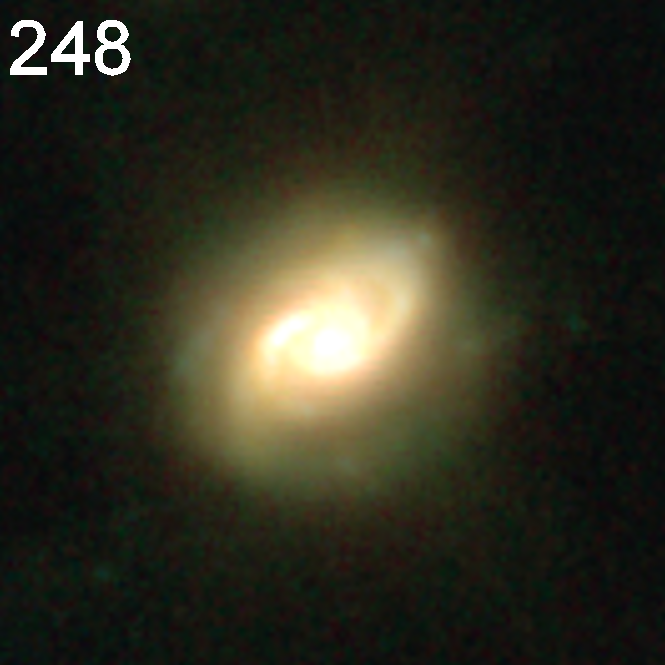}
\hfill
\includegraphics[width=0.495\linewidth]{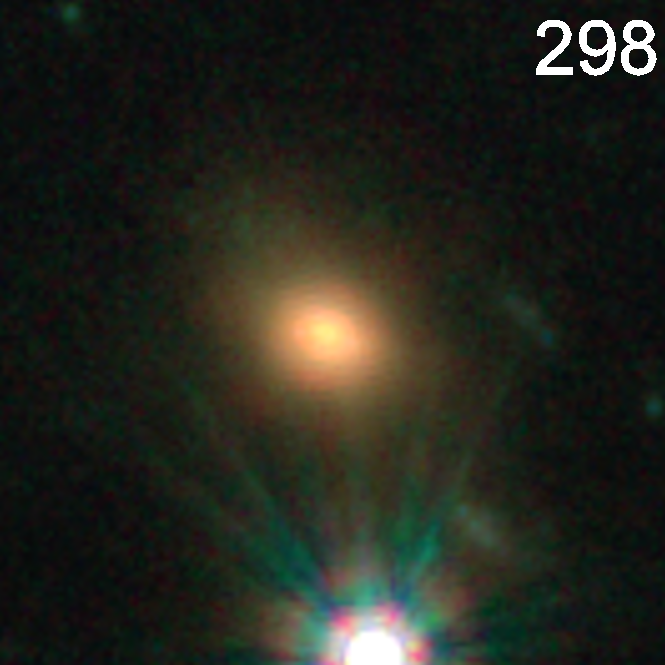}\\
\caption{Color images of four complex sources.  Sources are labeled, and each thumbnail is 4\arcsec\ square.  Colors are blue$=\text{F090W} + \text{F115W} + \text{F150W} + \text{F200W} + \text{F277W} + 0.8\times\text{F356W} + 0.5\times\text{F410M} + 0.5\times\text{F444W}$, green=$0.8\times\text{F200W}+\text{F277W} + 0.5\times\text{F356W}$, and red=$0.3\times\text{F356W} + 0.6\times\text{F410M} + \text{F444W}$.}
\label{f:color}
\end{figure}
}

Another possibility is that ID~193 is an infrared-faint radio source (IFRS; \citealt{Zinn2011}).  These are thought to be radio-loud AGN at high redshifts, perhaps $z>5$ \citep{Zinn2011,Herzog2016}.   In the local universe, galaxies less massive than the observational limit do not host luminous AGNs \citep{Sabater2019}, but the high-redshift population may include AGNs without much stellar mass around them. IFRSs have steep radio spectra, so ID~193 would have 1.4\,GHz flux density $>$46\,\mmJy, and the ratio of radio to NIR flux density $r\ga2000$.  For a source spectral energy distribution (SED) matching that of 3C~48, this ratio implies $z>3$ \citep[][their Fig.~4]{Zinn2011}.  Depending on the redshift, the source's radio luminosity would be 1--3 orders of magnitude smaller than that of 3C~48.  At ID~193's radio flux density, the IFRS number density is $\sim$16\,deg$^{-2}$, and the chance of having one such source in our field is $\sim$6\%.  The low likelihood, the source's proximity to a bright Seyfert galaxy, and the source being  extended make the IFRS hypothesis less attractive, but it cannot be ruled out.  ID~329 could also be an IFRS.  Its smaller flux density ($S\rm(20\,cm)\ga12$\,\mmJy) increases the space density, but the small offset from the galaxy nucleus argues for association with the galaxy.

\subsection{Reliability}
\label{s:reliability}

Figure~\ref{f:xy} shows the coordinate offsets from the radio positions to the positions of the identified counterparts.  The median offset for the 63 identified sources is 0\farcs08, and the systematic offset between radio and NIRCam coordinates is $\le$0\farcs01 in both right ascension and declination.
These small offsets are the result of three factors. (1) the 0\farcs7 FWHM VLA 3\,GHz beam {provides} position uncertainties of 0\farcs1--0\farcs2 {even} for the faintest sources \citep{Hyun2023}. (2) Microjansky radio sources have median angular sizes  $<$1\arcsec\ FHWM \citep{Windhorst2003,Smolcic2017data,Cotton2018}, and there are no extended sources with much larger position errors in our sample. (Deconvolved radio source sizes in the sample are all $<$1\farcs6.) (3) The NIRCam positions are on the Gaia DR3 system and are accurate to $\lesssim$0\farcs03~rms, even for the faintest objects \citep{Windhorst2023}. As a consequence, the NIRCam--VLA offsets of our IDs are driven by the individual VLA position errors of 0\farcs1--0\farcs2 \citep{Hyun2023}. The small coordinate offsets verify the reliability of most of the identifications.

Figure~\ref{f:fz} shows the F444W magnitudes of the radio host galaxies \edit1{as a function of redshift}.  Nearly all are brighter than 23~mag and would be detectable in less sensitive 4.4--4.5\,\micron\ surveys. Three counterparts are fainter than that with the faintest at 24.13~mag. In other words, the JWST detections of \edit1{compact} \mmJy\ radio sources are  easy for NIRCam, about 4~mag above the F444W point-source 5$\sigma$ detection limit \citep[28.35\,AB:][]{Windhorst2023}.  The galaxy counts rise steeply to $\rm AB\lesssim29$~mag \citep{Windhorst2023}, and spurious identifications unrelated to the radio source would nearly all be near the image detection limit, contrary to what Figure~\ref{f:fz} shows.  The actual detections being far above the limit is strong evidence that the contamination fraction among our current sample of VLA IDs must be small.   The median F444W magnitude of the radio counterparts is 20.9~mag. For comparison, the typical $r$-band magnitude  of \mmJy\ samples  peaks around  $r\sim22$~mag, as summarized by, \eg, \citet[][their Fig. 4]{Windhorst2003} and \citet{Smolcic2017id}.

\subsection{Special sources}
\label{s:special}

Some tricky  or interesting sources  shown in Figure~\ref{f:bigsep} or~\ref{f:tricky} are described below by ID. \edit1{Figure~\ref{f:color} shows color images of four complex sources.}
\begin{itemize}
\item[194:] this is the bright, extended Seyfert galaxy discussed in Section~\ref{s:missing}.  Its central region is saturated at wavelengths $>$3~\micron, and the SExtractor F444W position refers to the centroid of the extended galaxy body.  Measured at shorter wavelengths, the nucleus is 0\farcs03 from the radio position as opposed to 0\farcs47 indicated by the automated search.

\item[203:] the automated search found three sources brighter than the correct counterpart within a 1\arcsec\ radius but did not find the counterpart.  The \edit1{nearest-neighbor} match is 0\farcs30 from the radio position and 2.8~mag brighter than the correct counterpart, which is $<$0\farcs03 from the radio position. This object is within 0.2~mag of being the faintest counterpart found.
The F444W image shows four clumps in roughly a N--S line, and the SW images show that the northernmost clump is double.  Numbering the major clumps C1--C4 from north to south, C1/2 are blue while C3/4 are red.  C4 is surrounded by many minor clumps, and it appears to be the nucleus of a clumpy, irregular galaxy.  The radio position corresponds to C3, which could be a clump within the irregular galaxy.  However, C3 is redder and brighter than any of the minor clumps. It may be a background source but is perhaps more likely a dusty star-forming galaxy (DSFG), which JWST is beginning to uncover in other contexts \citep[e.g.,][]{Barrufet2023,Bisigello2023,Cheng2023,Magnelli2023,Perez-Gonzalez2023}.

\item[213:] there is a large spiral galaxy about 1\farcs1 NW of the radio position, but the radio counterpart is a red, elongated object that is barely noticeable in the F200W image but prominent in F444W. The spectroscopic redshift 0.3608 likely refers to the spiral because Binospec would not have been able to observe the red source.

\item[223:] this source is pointlike and red and therefore presumably a quasar.

\item[248:] the automated search found a bright spiral galaxy 0\farcs27 from the radio position, but the radio counterpart looks to be a bright, blue clump in a spiral arm.  The clump's blue color and extension along the arm suggest it is more likely part of the spiral galaxy than a background source.   
The SW images resolve the structure into clumps, consistent with the radio source being a giant star-formation region in the spiral galaxy rather than a background source. Tables~\ref{t:ID} and~\ref{t:SEDfits} treat the galaxy as the counterpart, in particular to derive a photometric redshift.

\item[260:] the host galaxy is an elliptical with a bright, compact nucleus.  The spectrum is consistent with an early-type galaxy but shows [\ion{Ne}{5}], an AGN indicator \citep[e.g.,][]{Gilli2010,Mignoli2013}, in emission.  The radio emission is two orders of magnitude more than expected from star formation (Section~\ref{s:properties}) and presumably arises from the AGN.

\item[268:] this is a remarkable object: two elongated galaxies superposed one on another. A dark lane running southeast--northwest suggests the larger galaxy oriented in that direction is the one in the foreground.  SExtractor found only a blended source, but the foreground galaxy is by far the dominant contributor to the source flux, size, and orientation.  However, the position and east--west orientation of the radio source suggest the background galaxy (also oriented east--west) is the counterpart. The automated search gave only 0\farcs25 offset from the radio source, and therefore an automated search only slightly looser than 0\farcs24 would have identified the blended source as the counterpart with no indication of complexity.  

\item[291:] the host galaxy is an elongated disk with a compact nucleus.  The radio centroid is 0\farcs23 from the nucleus corresponding to a projected separation of $\sim$1.7~kpc at the galaxy's photometric redshift (Section~\ref{s:fits}). 
Some of the NIRCam images, especially F115W and F150W, show faint excess emission near the radio-source location.  The morphology suggests that much of the radio flux comes from a giant \hii\ complex located away from the galaxy nucleus.

\item[298:] this source has an offset of 0\farcs30, but there is only one plausible counterpart on any of the NIRCam images. Here an automated search would have found no counterpart at all unless the search radius was $\ge$0\farcs30, but a radius that large would have identified some incorrect counterparts, as seen above.  The radio source has a bright spot coincident with the infrared source and an extension or a separate spot offset by 0\farcs6 to the east.  The extension looks to be a radio lobe, which offsets the radio centroid from the nucleus.
The projected separation is $\sim$5\,kpc at the photometric-redshift (Section~\ref{s:fits}) distance. The brighter northwest component coincides with the host galaxy position, suggesting that the southwest component is a radio jet.  The host galaxy has a red, compact nucleus, consistent with an AGN.

\item[329:] this source has an offset of 0\farcs28 from the only plausible counterpart.  The radio source is just above a 5$\sigma$ detection, and its positional uncertainty is 0\farcs07, so this may be a case where random errors have conspired to make the offset larger than the 0\farcs24 threshold.  Alternatively, the radio source could be a lobe or other emitter offset from the galaxy nucleus, or the true counterpart could be an unrelated object not detected by NIRCam.  There is no trace of any such object in the images.

\item[363:] The host galaxy is a face-on spiral with a bright compact nucleus.  The spectrum shows strong, broad \ion{Mg}{2} in emission and possibly weak \ion{C}{2} as well.  These indicate a Type~1 AGN, which may contribute much of the radio emission.

\end{itemize}

\section{Analysis} \label{s:disc}
\subsection{SED fitting method}
\label{s:fits}

\begin{figure}
\centering
\includegraphics[width=0.9\linewidth, clip=true, trim=36 70 80 24]
{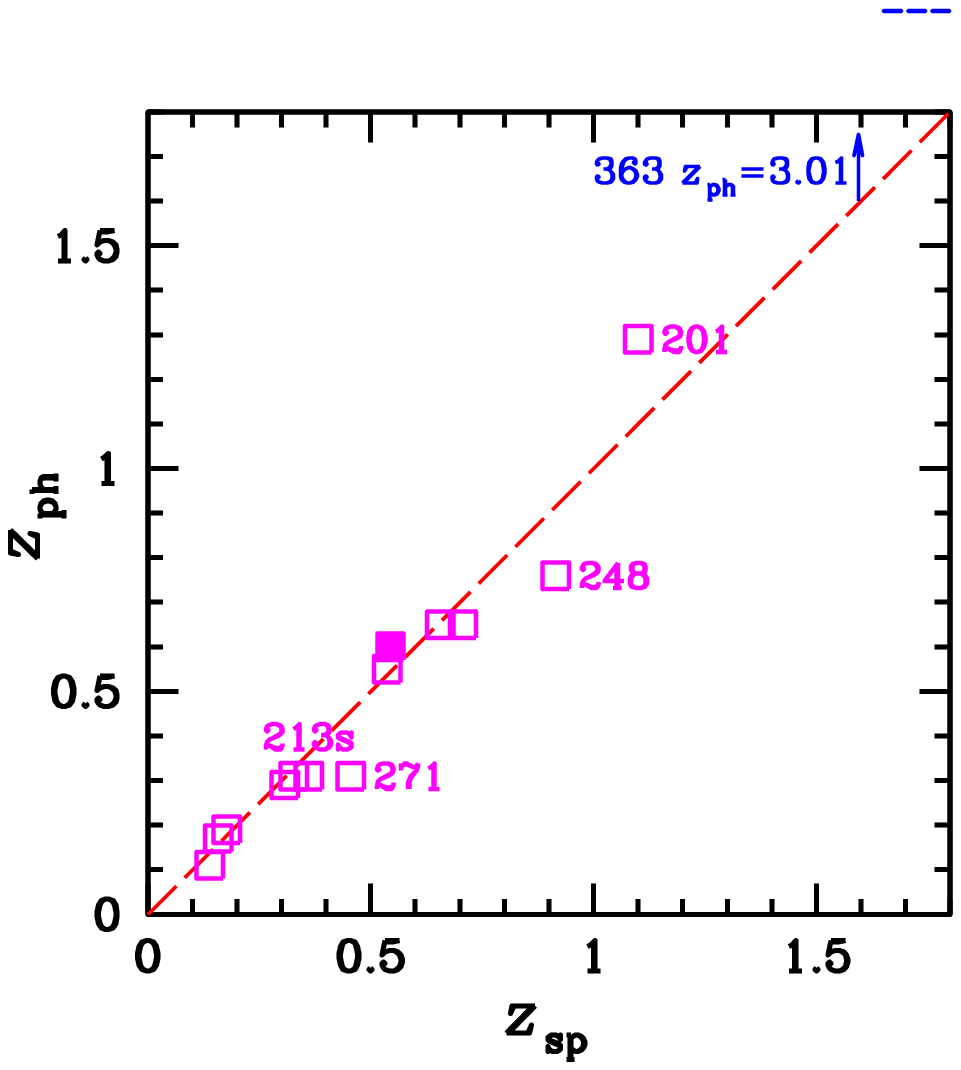}
\caption{Photometric versus spectroscopic redshifts for 14 sources with both redshift types available. Photometric redshifts are from CIGALE as discussed in Section~\ref{s:fits}.
Magenta squares represent galaxies with SED type (Section~\ref{s:nature}) Gal.
The filled square represents ID~260, which has excess radio emission as defined in Section~\ref{s:properties}. Two points \edit1{(one at $z\approx0.18$ and the other at $z\approx0.33$)} are shifted in both axes by $\Delta z=-0.02$ to separate them from nearby points.
The diagonal line shows equality, and the arrow indicates the one catastrophic failure, ID~363, which is SED type QSO.  Points with $|\zp-\zsp|>0.1$ are labeled. The spiral galaxy near ID~213 (labeled ``213s'') is included because it is a test of \zp\ even though it is not the radio-galaxy host.
 }
\label{f:zz}
\end{figure}

To derive host-galaxy properties, we used photometry from NIRCam \edit1{(Section~\ref{s:jwst})} and from HSC and MMIRS \edit1{(Section~\ref{s:ground})}.  Most NIRCam flux densities came from applying SExtractor to each  image in dual-image mode, using F444W as the detection image. For ID~244/270, which are not on the F444W image, F200W was used instead. This automatically measures flux densities in the same area on each image.  Colors were derived from {\sc mag\_iso} differences and applied to {\sc mag\_auto} from the F444W image.  For HSC and MMT magnitudes, counterparts were matched to sources in the \citet{Willmer2023} HSC--MMIRS catalog based on positions.  In practice, the matching was unambiguous, but some NIRCam sources lack ground-based counterparts because the JWST data are deeper, and many sources are red.
For most sources in the HSC--MMIRS catalog, ``auto'' magnitudes at \textit{iJHK} and in the NIRCam F090W--F277W filters fit a consistent SED. However, for ID~223, we needed aperture photometry in a 2\farcs0 aperture to make the HSC and MMIRS photometry agree with the NIRCam photometry. ID~298 is too red to be detected in the ground-based data, and we used only the NIRCam data for the SED fitting.

Spectral energy distributions (SEDs) were analyzed with CIGALE \citep{Boquien2019, Yang2020, Yang2022}.\footnote{Computation was performed on the Tempest High Performance Computing System, operated and supported by University Information Technology Research Cyberinfrastructure at Montana State University.} For each source, CIGALE fit a suite of galaxy models to the observed photometry at  redshifts ranging from 0.01 to \edit1{8}, evenly spaced on a logarithmic scale.  The galaxy models included a wide range of stellar, dust, and AGN parameters, described below, and we adopted final parameters from the model with smallest $\chi^2$. Of the 62 identified radio-source host galaxies (ID~193 having no host identified), 59 have reduced $\chi^2_r< 2.5$, and their parameters \edit1{along with estimates of uncertainty ranges from the respective probability density functions (PDFs)} are listed in Table~\ref{t:SEDfits}.
Figure~\ref{f:zz} shows good agreement between photometric and spectroscopic redshifts except for one catastrophic failure (ID~363).  Excluding that, the mean $|\Delta z|/(1+\zsp)=0.026$.
As mentioned in Section~\ref{s:radio}, the field is near a \edit1{$z=1.44$} quasar. Five sources have $1.36<z\le1.52$, and their projected separations all correspond to $<$1.7\,Mpc at $z=1.44$. This could indicate an overdensity marked by the quasar.
  
For stellar parameters, the suite of models included three star-formation histories (SFH): double exponential (refer\-red to as \textsl{sfh2exp}), delayed SFH with optional burst (\textsl{sfhdelayed}), and delayed SFH with optional burst and quenching (\textsl{sfhdelayedbq}). The main ages for all SFH models were $\tau_{\rm main} =0.1$, 0.5, 1, 2.5, 5~Gyr and  $\rm age_{main} =0.5$, 1, 3, 5, 7, 9~Gyr.
For the \textsl{sfhdelayed} model, trial values of the mass fraction of the late-burst population were 0.0 and 0.001 with a constant $e$-folding time of the late-burst population model at 50~Myr and an age of the late burst of 20~Myr. The \textsl{sfhdelayedbq} model included additional parameters for the burst/quench episode ages of 100 and 500~Myr and a ratio of star formation after/before the burst/quench age of 0.1.
For the simple stellar population, we utilized  models by \citet{Bruzual2003} with Salpeter initial mass function \citep{Salpeter1955} and metallicities of 0.0001, 0.004, and 0.02, i.e., 1/200, 1/5, and 1 times solar. The dust attenuation model assumed the \citet{Calzetti2000} prescription with $E(B-V)=  0.01$, 0.1, 0.25, 0.5, 1.0, and 1.5 for emission lines and, for the stellar continuum, either the same as or half of the emission-line value.

For the dust emission, the model suite included the \citet{Draine2014} model and the Themis model \citep{Jones2017}. Trial values for the mass fraction of polycyclic aromatic hydrocarbons (PAH) for the \citeauthor{Draine2014} model were 2.5 and 5.26 and for the Themis model were 0.14 and 0.28. For both models, the dust mass $M_d$ at each value of radiation field intensity $U$ was taken to be a power law $dM_{d} (U)/dU \propto U^{-\beta}$ with $\beta= 1.5$ and 2.5. CIGALE incorporates two emission models, diffuse emission and clumpy emission, controlled by the parameter $\gamma$. Values  $0.01\le\gamma\le0.99$ had negligible impact on the results within our wavelength range, and therefore we set  $\gamma=0.1$, the default value.

The AGN model was the Skirtor model \citep{Stalevski2012,Stalevski2016}.  We tested AGN fractions of 0.0, 0.3, 0.6, and 0.9 at edge-on optical depths at 9.7\,\micron\ of 3, 7, and 11. Viewing angles included were 30 degrees (representing type~1 AGN) and 70 degrees (representing type~2 AGN). 
All other AGN model parameters were kept at their default settings.
This set of values is sufficient to cover the likely ranges of AGN parameters. Differences of one step, e.g., between AGN fractions of 0.3 and 0.6, do not significantly change other derived parameters.

The three host galaxies without acceptable fits are ID~213, which has no useful photometry because of the superposed spiral galaxy; ID~244, which has useful photometry in only two SW filters; and ID~270, where there is no LW photometry, and the SW photometry is contaminated by diffraction spikes.  The fit for ID~248 is for the whole galaxy, not the blue clump (Section~\ref{s:special}) alone. The fit for ID~268 may not be meaningful because the photometry includes the \edit1{combined flux of the} two superposed galaxies, which may have different properties and different redshifts.  Because the fit for ID~298 used only NIRCam data, the resulting parameters are uncertain.  \edit1{The individual clumps in ID~203 have the same problem.  They are not resolved except by NIRCam, and the resulting photometric redshifts are uncertain.  The fits do not rule out all clumps having $\zph\approx4$ (Table~\ref{t:SEDfits}), but the color differences (Section~\ref{s:special}) suggest differing redshifts. For ID~223, the fit gives $f_{\rm AGN}=0$ despite the red point-source appearance, and the redshift is therefore dubious.  There are two wisps north of the point source, but we could not determine reliable \zph\ for those.}  For ID~363, the spectroscopic redshifts from Binospec and Hectospec agree, but the photometric redshift is a catastrophic failure (Figure~\ref{f:zz}).
\edit1{In general, fits for objects with high $f_{\rm AGN}$, as for ID~363, are less trustworthy than fits with low $f_{\rm AGN}$.}

\subsection{Source properties}
\label{s:properties}

\begin{figure}
\includegraphics[width=\linewidth, clip=true, trim=28 274 80 36]
{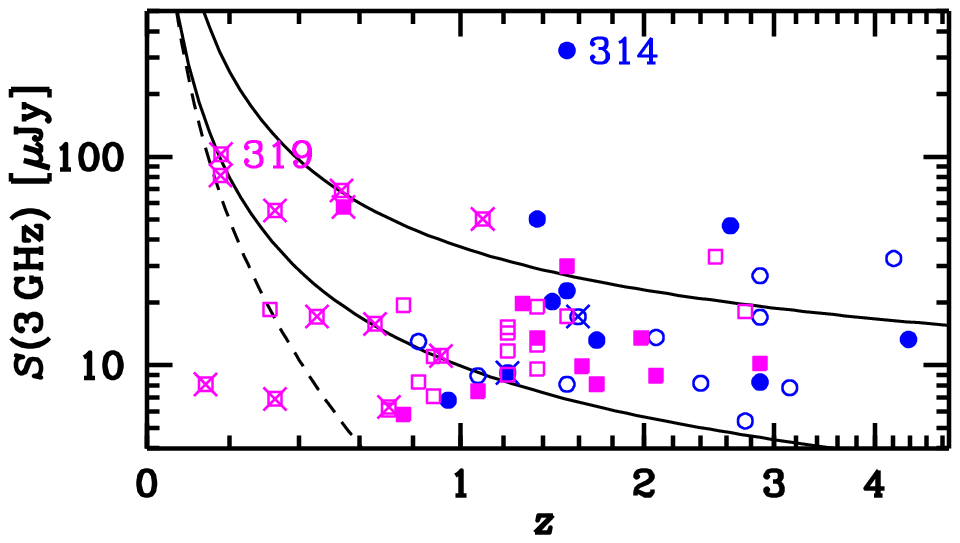}
\caption{3~GHz flux density versus redshift for 59 radio hosts that have redshifts measured. 
Point shapes and colors indicate SED classes defined in Section~\ref{s:nature}: blue circles represent type QSO, and magenta squares represent type Gal.  Filled symbols indicate galaxies with excess radio emission as defined in Section~\ref{s:properties}, and $\times$ symbols indicate galaxies with spectroscopic redshifts. 
The two points with highest flux density are labeled.  The solid lines show the flux density from star formation \citep[][corrected to a Salpeter IMF]{Murphy2011} for a galaxy on the main sequence \citep{Speagle2014} with stellar masses $10^{11}$\,\Msol\ (upper curve) and $10^{10}$\,\Msol\ (lower curve).  The dashed line shows the flux density for a fixed 1.4~GHz
luminosity density of $10^{22}$~W~Hz$^{-1}$, which corresponds to a star-formation rate of about 5~\Msol~yr$^{-1}$ \edit1{\citep[][their Eq.~3]{Mahajan2019}}.}
\label{f:sz}
\end{figure}

\begin{figure}
\vspace{-33pt}
\includegraphics[width=\linewidth,clip=true,trim=28 154 80 36]
{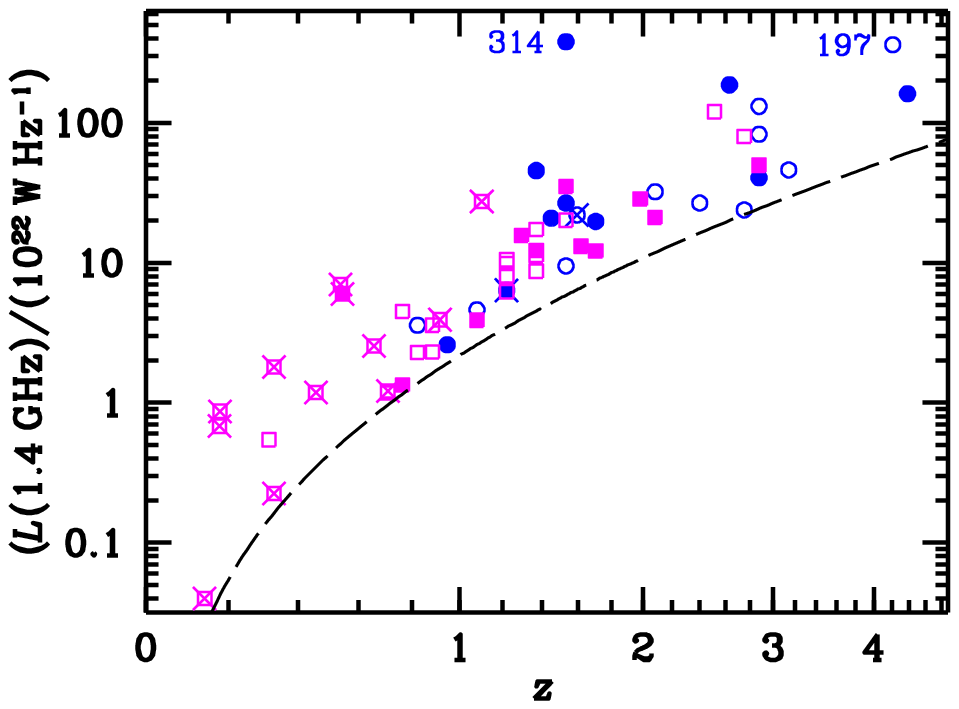}
\caption{Source luminosities versus redshift scaled as $\log(1+z)$. Luminosities were scaled from the observed 3\,GHz flux densities assuming spectral index $\alpha=-0.7$. Point shapes and colors indicate SED classes defined in Section~\ref{s:nature}: blue circles represent type QSO, and magenta squares represent type Gal.  Filled symbols indicate galaxies with excess radio emission as defined in Section~\ref{s:properties}, and $\times$ symbols indicate galaxies with spectroscopic redshifts. The two highest-luminosity hosts are labeled with their source IDs.  The dashed line shows the luminosity corresponding to 5\,\mmJy, the approximate radio survey limit.}
\label{f:Lz}
\end{figure}

The TDF radio sample spans a wide redshift range, 0.14 to 4.4 with a median  of 1.33.  
The wide range is 
consistent with previous results \citep[e.g.,][the COSMOS-XS survey]{Algera2020}: radio flux densities have little or no dependence on redshift (Figure~\ref{f:sz}). Figure~\ref{f:Lz} is a direct analog of \citeauthor{Algera2020}'s Figure~4b and is consistent with it given our shallower radio sensitivity and smaller survey area.
The COSMOS-XS's median redshift is 1.0, a bit smaller than found here, but  the difference is not physically significant considering our small sample size and the different input data and methods for photometric redshifts.

\begin{figure}
\includegraphics[width=\linewidth, clip=true, trim=28 124 72 72]
{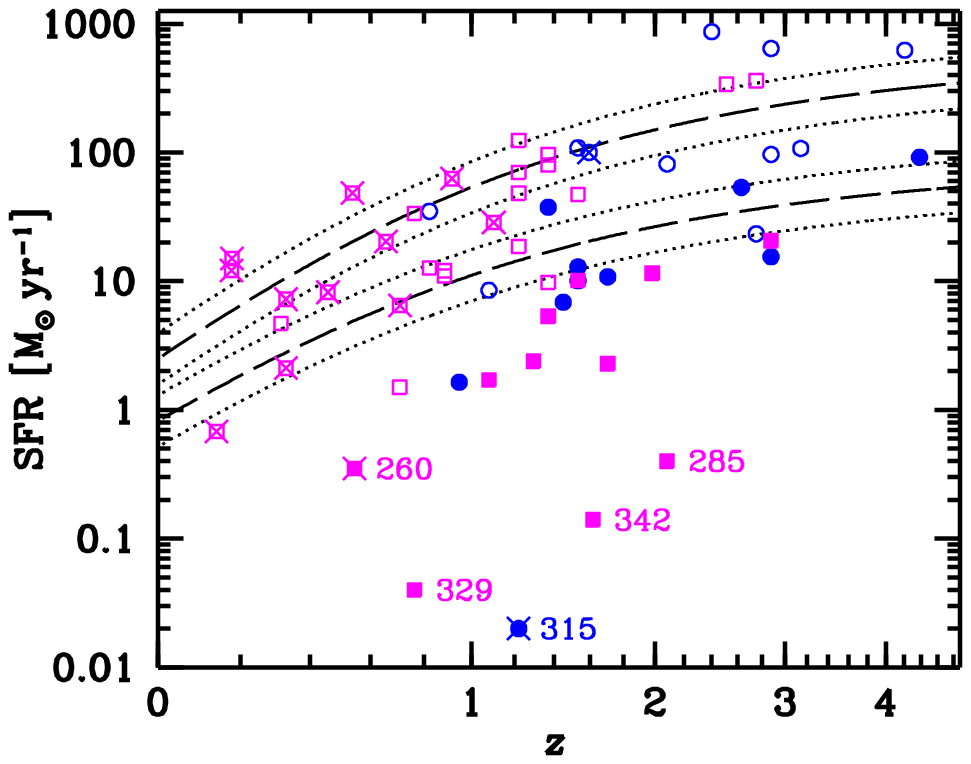}
\caption{Photometric star-formation rate versus redshift  scaled as $\log(1+z)$.  Point shapes and colors indicate SED classes defined in Section~\ref{s:nature}: blue circles represent type QSO, and magenta squares represent type Gal.  Filled symbols indicate galaxies with excess radio emission as defined in Section~\ref{s:properties}, and $\times$ symbols indicate galaxies with spectroscopic redshifts. Long-dashed lines show the galaxy main sequence \citep{Speagle2014} for stellar masses of $10^{10}$ and $10^{11}$\,\Msol, and dotted lines show intervals of 0.2\,dex on either side.  SFRs were derived from the CIGALE fits (Section~\ref{s:fits}) to the NIR photometry and assume a Salpeter IMF.}
\label{f:z_sfr}
\end{figure}

\begin{figure}
\includegraphics[width=\linewidth]{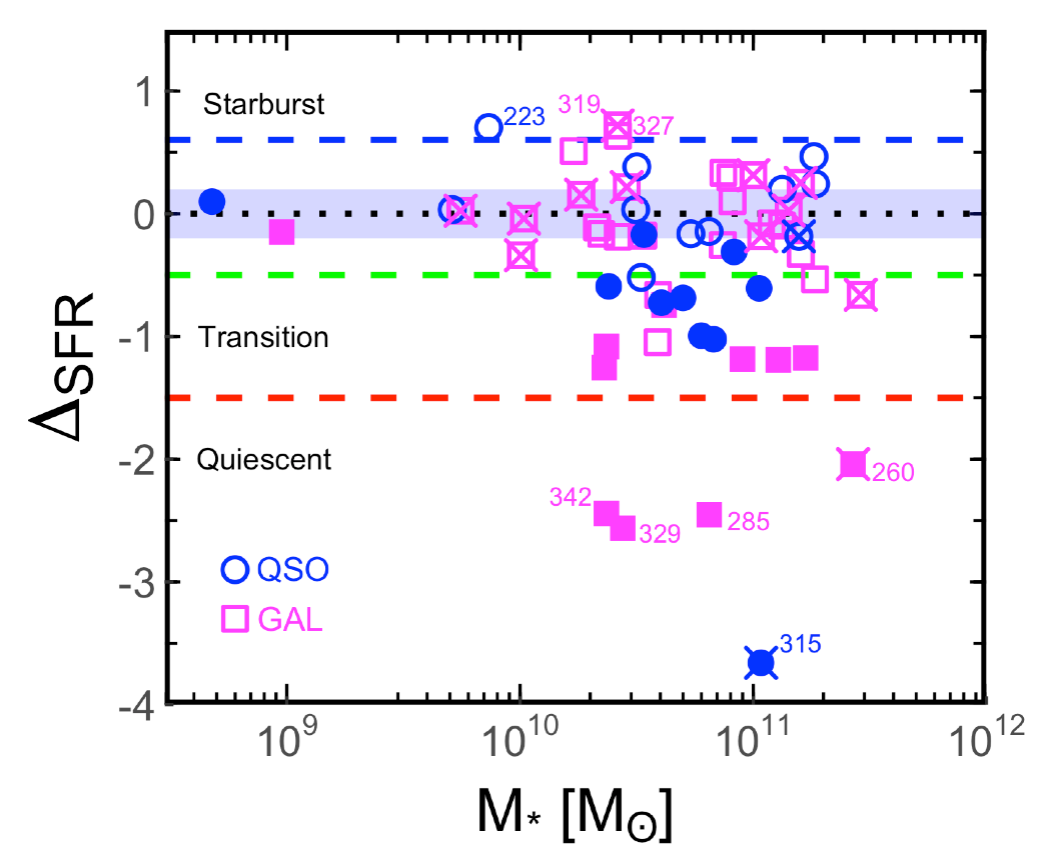}
\caption{Star-formation rates from the CIGALE fits relative to the star-formation main sequence.  The vertical axis shows the difference $\rm \Delta_{SFR}\equiv\log_{10}(SFR)/\log_{10}(SFR_{MS})$ with $\rm SFR_{MS}$ as defined by \citet{Speagle2014} for the mass and redshift of each galaxy but scaled \citep{Madau2014} to a \citet{Salpeter1955} IMF.  
The horizontal dotted line marks the main sequence, and the shaded region shows the $\pm$0.2\,dex range around it.  The upper dashed line marks 0.6\,dex above the main sequence, commonly taken as the boundary for a starburst \citep[e.g.,][]{Rodighiero2011}, and the other two dashed lines mark the lower boundary of the main sequence and the upper boundary of the quiescent region \citep[$\Delta_{\rm SFR}=-0.5$ and $-1.5$, respectively:][]{Renzini2015}.
Point shapes and colors indicate SED classes defined in Section~\ref{s:nature}: blue circles represent type QSO, and magenta squares represent type Gal.  Filled symbols indicate galaxies with excess radio emission as defined in Section~\ref{s:properties}, and $\times$ symbols indicate galaxies with spectroscopic redshifts.  }
\label{f:delsfr}
\end{figure}

The star formation rates (SFRs) from the SED fitting (Section~\ref{s:fits}) range from \edit1{$\sim$0.02} to $\sim$900\,\Msol\,yr$^{-1}$ as shown in Figure~\ref{f:z_sfr}, although the highest SFRs could be much lower and still fit the observed photometry. 
Figure~\ref{f:delsfr} compares the SFRs to the star-formation main sequence \citep{Speagle2014}. Three host galaxies (5\% of the sample) are starbursts, 34 (58\%) are normal star-forming galaxies with $\rm -0.5 \le \Delta_{SFR} < 0.6 $,  17 (29\%) are galaxies in transition with $\rm -1.5 \le \Delta_{SFR} < -0.5$, and 5 (8\%) are quiescent with $\rm \Delta_{SFR} < -1.5$. 
(Range definitions are from \citealt{Rodighiero2011} and \citealt{Renzini2015}.)

\subsection{Nature of the sources}
\label{s:nature}

\begin{figure}
\includegraphics[width=\linewidth]{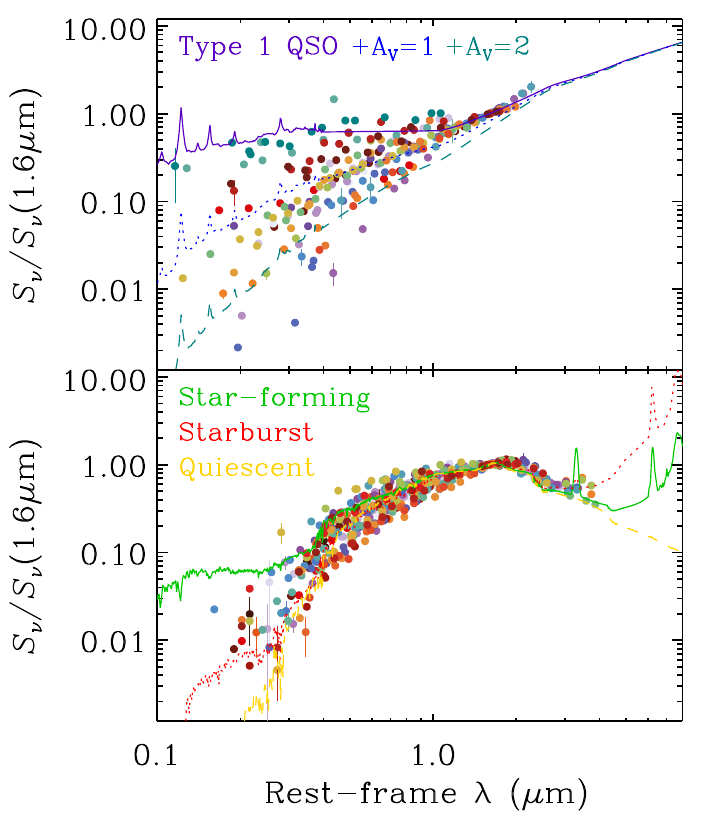}
\caption{Rest-frame SEDs of the radio-source host galaxies. Filled-circle colors identify individual host galaxies.  (Color choice is arbitrary.)  Panels show galaxies in each SED class: QSO top, Gal bottom. Flux densities are normalized at rest-frame 1.6\,$\mu$m.  Lines in each panel show templates from the SWIRE library \citep{Polletta2007}. The top panel has a type~1 QSO template,  TQSO1, in purple (solid line). The same template reddened with $A_{V}=1$ is in blue (dotted line) and with $A_{V}=2$ in teal (dashed line). The bottom panel shows templates for a star-forming galaxy in green (solid line, top in UV, middle in IR), a starburst galaxy in red (dotted line, middle in UV, top in IR), and an old stellar population in yellow (dashed line, bottom in both UV and IR).}
\label{f:rfseds}
\end{figure}

The source SED shapes fall into two broad categories:  SEDs that rise beyond rest-frame 1.6\,\micron, consistent with being AGN-dominated (labeled ``QSO''), 
and  stellar-dominated SEDs (labeled ``Gal'') with falling SEDs beyond rest 1.6\,\micron. 
This classification is based on the visible--NIR SED, where hot dust emission produces an SED  redder than that from a stellar population \citep{Sanders1989} and a dip at $\sim$1\,\micron. Such hot dust ($T_d\ga1000$\,K) is usually associated with an AGN. For power-law SEDs, we assumed that the AGN dominates also at rest-frame visible wavelengths and classified those objects as QSO. In some cases, the SED reddens at shorter wavelengths, consistent with a reddened QSO \citep{Gregg2002}. 
Figure~\ref{f:rfseds} shows the SEDs of  sources in the two categories, and
Table~\ref{t:classification} shows the number of sources in each category.
Overall, 36\% of radio host-galaxy SEDs show evidence for an AGN.

\begin{figure}
\includegraphics[width=\linewidth]{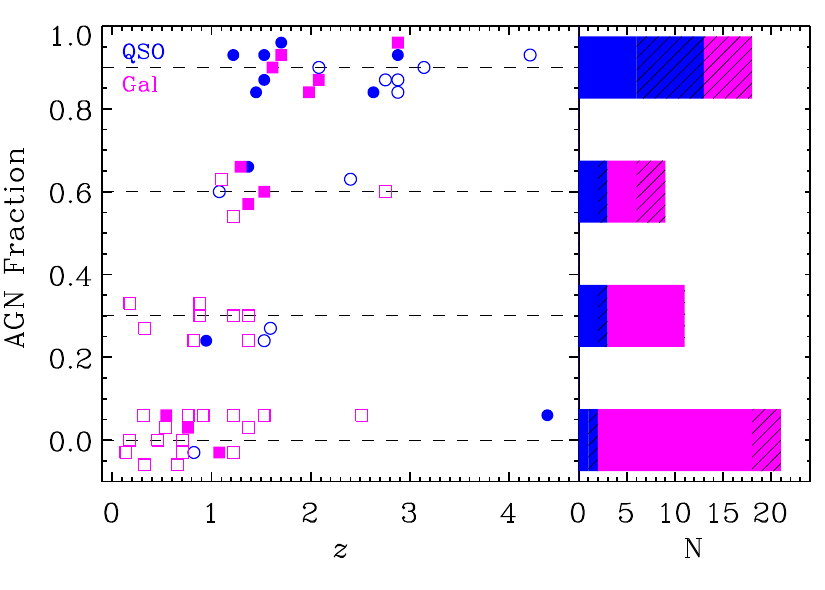}
\caption{Comparison between CIGALE AGN fraction and the SED class. Sources classified QSO are shown as blue circles, and Gal as magenta squares. Filled symbols represent sources with a radio excess. The AGN fractions have been slightly offset from their values of 0, 0.3, 0.6, and 0.9 (horizontal dashed lines) for visualization. The right panel shows the number of sources of each SED class, coded with the preceding colors, for each level of AGN fraction. Hatched regions refer to sources with a radio excess. }
\label{f:agn_fraction}
\end{figure}

The SED classification will not identify an AGN whose emission does not contribute to visible--NIR wavelengths. This might be the case for a massive host galaxy (\eg, some powerful radio galaxies---\citealt{Haas2008}) or when dust obscures the AGN at rest visible wavelengths which are redshifted to the NIRCam range, \ie, at $z\gtrsim2$, the dust would emit at $\lambda>5$\,\micron, where there are no available data. 
To assess the classification's reliability, Figure~\ref{f:agn_fraction} compares the CIGALE AGN fraction with the SED classification. The SED classification and CIGALE agree for 64\% of the sample. 90\% of the QSO have an AGN fraction $\geq$30\%, and 50\% of the Gal have a null AGN fraction. 90\% of the sources with $f_{\rm AGN}=0$ are classified as Gal, and 72\% of those with $f_{\rm AGN}=0.9$ are classified QSO. Of sources classified as Gal, 37\% have $f_{\rm AGN}=0.3$ or 0.6. In the following, we will use the SED classification to identify AGNs. This is because CIGALE sometimes includes an AGN component that emits at wavelengths where there are no data, and it never produces a model where the AGN dominates at visible wavelengths, contrary to what is observed in QSOs. On the other hand, based on the comparison with CIGALE, we should assume that the SED classification might be incorrect in about 20\% of the cases. More extensive wavelength coverage combined with morphological analysis might provide more reliable AGN identification.

\begin{figure}
\includegraphics[width=\linewidth]
{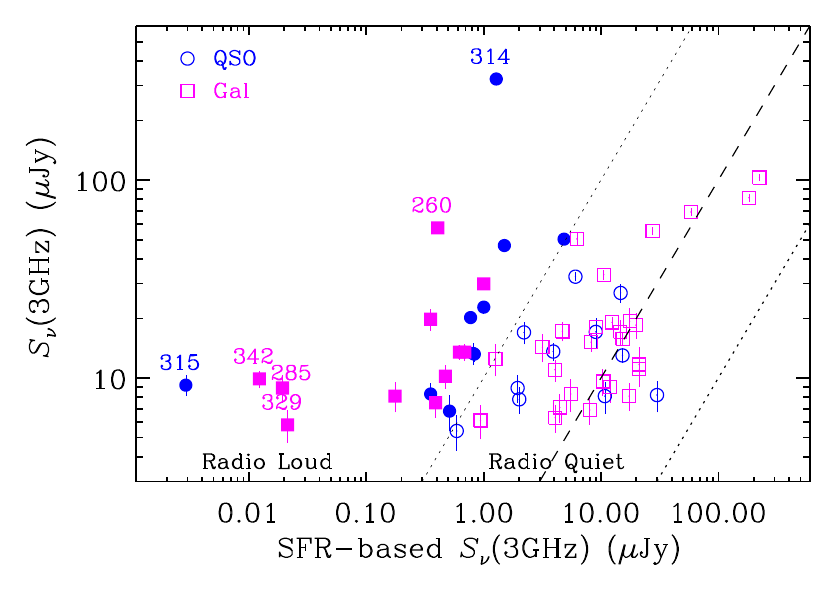}
\caption{Comparison of expected to observed radio flux density.  The ordinate shows the observed 3\,GHz flux density, and the abscissa shows the flux density expected from the SFR derived from the CIGALE SED fitting.  The predicted radio emission for a given SFR was taken from \citet{Murphy2011}, and the radio spectral index was assumed to be $-0.8$.  Lines show  equality and 1\,dex scatter. Symbols as in Figure \ref{f:agn_fraction}.}
\label{f:S3_sfr}
\end{figure}

For each source, the SFR and \zp\ derived from CIGALE predict the rest 1.4\,GHz radio flux density \citep{Murphy2011}. Figure~\ref{f:S3_sfr} compares these predictions to the observed flux densities.  Out of the 59 radio sources with an SFR estimate, 38 (64\%) exhibit radio emission  within a factor of ten of the value predicted from the SFR. Given the uncertainties in the derived SFR, this means no strong need to invoke any process other than star formation to produce the radio emission.  The remaining 21 sources (36\%) have excess radio emission  with respect to the value derived from the SFR. For present purposes, we assume this emission is from an AGN, but it could instead arise from star formation hidden behind dust and not contributing to the observed SED.  The most extreme source in terms of $\Delta_{\rm SFR}$ is a reddened QSO at $\zph=1.22$ (ID~315), which not surprisingly has radio emission above that expected from its SFR. The four other quiescent sources also show AGN-powered radio emission, consistent with being radio galaxies, and three of the four are classified Gal (Figure~\ref{f:delsfr}). These are candidates to have hidden star formation.

AGN selection should include both the SED and radio excess as indicators.  Radio excess adds eleven Gals to the AGN list and raises the AGN fraction from 36\% to 54\%. This fraction is higher than the VLA-COSMOS survey \citep{Smolcic2017id}, where 40\% of the $S(3\,\rm GHz) \sim 50$\,\mmJy\ radio population are AGNs but could be consistent (Poisson probability 7\%) given the small number of sources in our sample.  A more intriguing explanation is that much of the UV emission from young stars is hidden behind dust and not found by CIGALE. Larger samples will be needed to explore this possibility.  Regardless of AGN evidence, radio emission comes principally from star formation in 64\% of sample galaxies.

\begin{table}
\caption{Source SED classifications\label{t:classification}}
\begin{tabular}{l r@{\quad\quad} c c}
\hline
Class   & $N$    &  \multicolumn{2}{c}{Source of radio emission} \\
        &        &      AGN   &    Star Formation    \\
\hline
QSO     & 21     &    10      &    11     \\
Gal     & 38     &    11      &    27     \\
\hline
\end{tabular}
\end{table}

\begin{figure}
\includegraphics[width=\linewidth]
{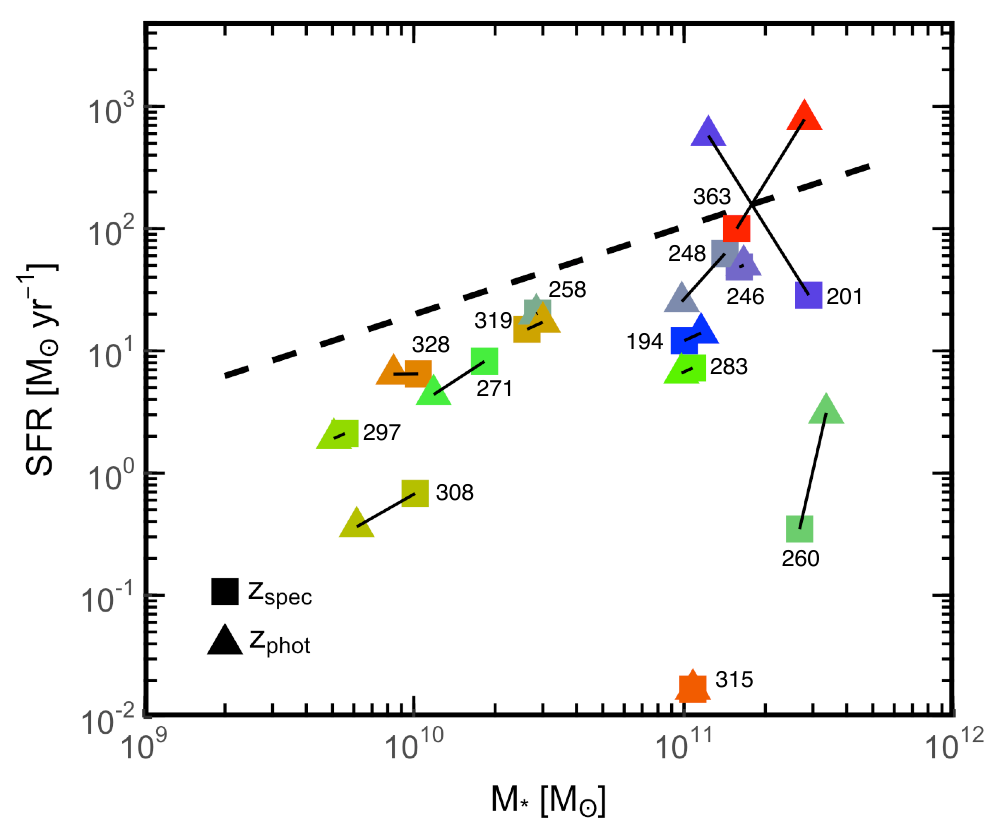}
\caption{Star-formation rate comparison for sources with spectroscopic redshifts. Triangles show \zph, and squares show \zsp\ with lines connecting the two for individual objects.  SFR, \zph, and $M_*$ are from the CIGALE fits.   The diagonal line shows the star-formation main sequence \citep{Speagle2014} at $z=1.332$.}
\label{f:sfrcomp}
\end{figure}

\begin{figure*}
\includegraphics[width=0.49\linewidth]{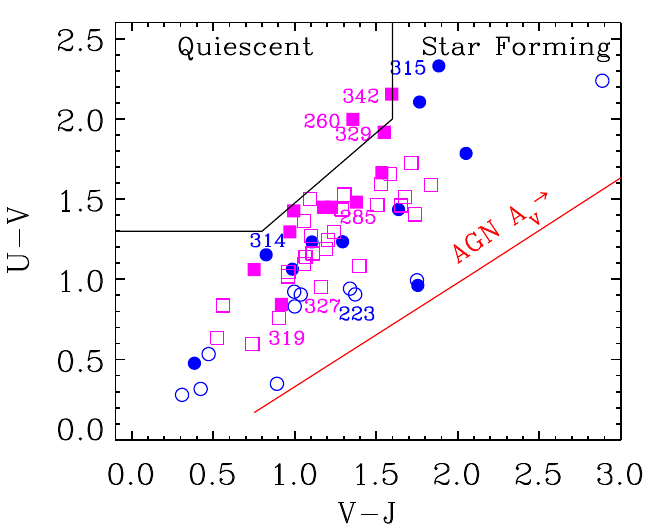}
\includegraphics[width=0.49\linewidth]{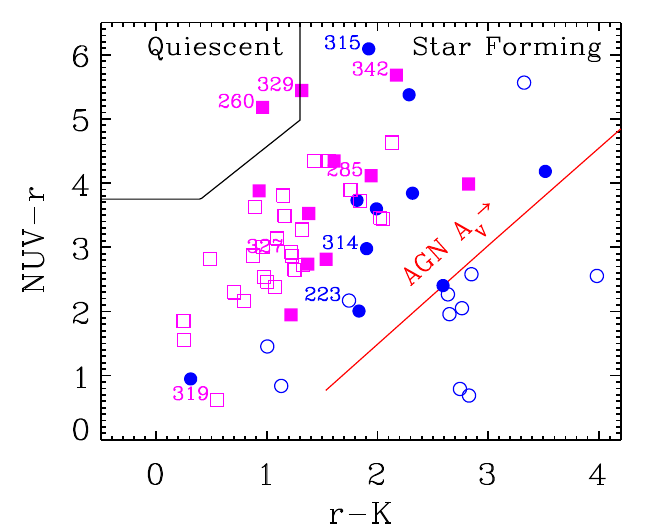}
\caption{Two color--color diagrams for identifying quiescent galaxies. Left panel is rest-frame \textit{U--V--J} \citep{Williams2009}, and right panel is rest-frame near-ultraviolet--$r$--$K$ \citep{Arnouts2013}. Marked regions at upper left of each panel show the colors of quiescent galaxies. Star-forming galaxies with increasing extinction lie on a sequence from the bottom left to the top right. The rest-frame colors of a type 1 QSO with increasing extinction levels (A$_{\rm V}$ increases from 0 in the bottom left to 4 in the top right) are shown by the solid red line. Symbols as in Figure \ref{f:agn_fraction}. Labeled sources are those classified starburst based on their offset from the main sequence (IDs 223, 319, and 327), as quiescent (IDs 260, 285, 315, 329, and 342), and those with a radio excess (IDs 260, 285, 314, 315, 329, and 342). }
\label{f:passive}
\end{figure*}

Two other indicators of whether a galaxy is star-forming or quiescent are the \textit{UVJ} and NUV--\textit{rK} color--color diagrams \citep{Williams2009,Arnouts2013}. Figure~\ref{f:passive} shows both diagrams with rest-frame colors computed by convolving the CIGALE best-fit, rest-frame model of each source with each filter's transmission curve. Of the five sources that should be quiescent based on $\Delta_{\rm SFR}$ (Figure~\ref{f:delsfr}), two  (ID~260/329) are in the quiescent wedge. Their radio emission is also higher than expected based on their SFRs (Figure~\ref{f:sfrcomp}), and they can be considered radio galaxies. Two other sources (ID~315/342) fall in the region of dusty, star-forming galaxies. Their radio fluxes indicate SFRs $\sim$3\,dex higher than the SED fits ($\sim$60--90\,\Msol\,yr$^{-1}$ vs.\ $\la$0.1), which would put these galaxies on the main sequence. Both galaxies exhibit extremely red visible SEDs that CIGALE fits with an old stellar population. Based on the radio fluxes, their red colors might be due to dust extinction instead. Whether these sources are quiescent with AGN-powered radio emission or are dusty, star-forming galaxies therefore remains in doubt. The last $\Delta_{\rm SFR}$-quiescent source (ID~285) is outside the color--color quiescent wedges and might be in a transition phase. This source also exhibits a radio excess, indicating that it hosts an AGN. Because of its $\zph\sim2$, the photometry does not sample rest-frame NIR emission, and therefore the SED classification cannot assess whether AGN hot dust is present. However, CIGALE prefers an AGN fraction of 90\%. Spatially resolved spectroscopy might show evidence of quenching due to AGN feedback. At the other end of SFR, the three starbursts (ID~223/319/327) identified by  $\Delta_{\rm SFR}$ fall in the blue star-forming galaxy region of the color--color diagrams, as expected.

\section{Summary and Conclusions}
\label{s:summ}

Even \edit1{$\sim$50~minute} exposure times with the JWST/\linebreak[0]NIRCam F444W filter can detect counterparts for at least 97\% of 5\,\mmJy\ radio sources and plausibly 100\%.  However, finding the correct counterpart for a given source can be tricky.  Simple position matching with separation $<$0\farcs24 would have been 100\% reliable in the current sample but would have missed 11\% of sources.  The existence of incorrect matches at 0\farcs25 separation suggests that there is nothing fundamental about a limit at 0\farcs24.
Indeed, even a very small match radius could give incorrect counterparts when sources overlap.  Had the spiral galaxy overlapping ID~213 happened to be close enough to the radio position, even JWST's superb angular resolution could not have found the correct counterpart.
At the other extreme, matching within a separation of 0\farcs5 would have found all counterparts but would have found incorrect ones for 11\% of sources.  No single match radius would give both completeness and reliability.  This is not because of position errors but because sources have intrinsic offsets between radio and infrared positions, for example because of jets or asymmetric structures or because of confusion by nearby or overlapping sources.

While the bulk of the radio host galaxies are the same population as seen in previous surveys, as they must be given previous identification rates, NIRCam identifies galaxies that would be difficult or impossible to have seen before.  The faintest 5\% of sources are red and have $\rm F444W>22$\,mag.  Two of these have QSO-type SEDs, but one is type Gal.  While missing these would not change the overall statistics of the sample, they may represent a different population.  Another new aspect is NIRCam's ability to identify confused or complex sources.  A good example, aside from the overlapping sources (IDs~213/268), is ID~291, where much of the radio flux appears to come from an \hii\ region. Another is ID~329, where the position offsets suggests jet emission, as is also plausible for ID~193.

At this writing, JWST has observed all four spokes of the TDF. The full image will give a sample four times larger than the one used here.  That will measure what fraction of the radio population is DSFGs.  The numbers may even be large enough to constrain the respective luminosity functions. 
Larger numbers will also allow better comparison of different methods to tell whether emission is from star formation or an AGN.  That may include use of morphological information, even pixel-level image decomposition to measure the sources' angular sizes, and comparison to radio sizes measured by very long baseline interferometry.
Larger numbers and NIRCam images in multiple directions around the quasar at the center of the radio field will show whether there is an overdensity around the quasar.
The parallel NIRISS data should give more spectroscopic redshifts, addressing the putative \edit1{$z=1.44$} overdensity and also providing SFR measurements from emission lines and additional AGN indicators from line ratios. \edit1{MIRI observations---not available now but planned to be proposed---would provide another AGN indicator.} The key for all of these studies will be identifying the correct counterparts.

\bigskip
%\begin{acknowledgements}
\edit1{This paper is dedicated to the memory of PEARLS team member
  and collaborator Mario Nonino, whose enthusiasm for the science and
  generosity have been an inspiration. We thank Denis Burgarella for
  his invaluable assistance with CIGALE.} 
This work is based on observations made with the NASA/ESA/CSA James Webb Space
Telescope. The data were obtained from the Mikulski Archive for Space
Telescopes at the Space Telescope Science Institute, which is operated by the
Association of Universities for Research in Astronomy, Inc., under NASA
contract NAS 5-03127 for JWST. These observations are associated with JWST
programs 1176 and 2738.
\edit2{Observations reported here were obtained at the MMT Observatory, a joint facility of the Smithsonian Institution and the University of Arizona.
This work is based in part on data collected at the Subaru Telescope and retrieved from the HSC data archive system, which is operated by the Subaru Telescope and Astronomy Data Center at the National Astronomical Observatory of Japan.}
SHC, RAJ, RAW, and HBH acknowledge support from NASA JWST Interdisciplinary
Scientist grants NNX14AN10G, 80NSSC18K0200,  NAG5-12460, and 21-SMDSS21-0013, respectively, from  NASA Goddard Space Flight Center (GSFC). 
CNAW acknowledges support from the NIRCam Development Contract NAS5-02105
from GSFC to the University of Arizona.
NJA and CJC acknowledge support from the European Research Council (ERC) Advanced
Investigator Grant EPOCHS (788113). BLF thanks the Berkeley Center for
Theoretical Physics for their hospitality during the writing of this paper.
MAM acknowledges the support of a National Research Council of Canada Plaskett
Fellowship, and the Australian Research Council Centre of Excellence for All
Sky Astrophysics in 3 Dimensions (ASTRO 3D), through project number CE17010001.
CNAW acknowledges funding from the JWST/NIRCam contract NASS-0215 to the
University of Arizona.
JFB was supported by NSF Grant No.\ PHY-2012955.
\edit1{KEN and WMP acknowledge that basic research in radio astronomy at the U.S.\ Naval Research Laboratory is supported by 6.1 Base funding.} 
M.H.\ acknowledges the support from the Korea Astronomy and Space Science Institute grant funded by the Korean government (MSIT) (No.\ 2022183005) and the support from the Global Ph.D.\ Fellowship Program through the National Research Foundation of Korea (NRF) funded by the Ministry of Education (NRF-2013H1A2A1033110).
We also acknowledge the indigenous peoples of Arizona, including the Akimel
O'odham (Pima) and Pee Posh (Maricopa) Indian Communities, whose care and
keeping of the land has enabled us to be at ASU's Tempe campus in the Salt
River Valley, where much of our work was conducted.

LOFAR data products were provided by the LOFAR Surveys Key Science project (LSKSP; \url{https://lofar-surveys.org/}) and were derived from observations with the International LOFAR Telescope (ILT). LOFAR \citep{lofar} is the Low Frequency Array designed and constructed by ASTRON. It has observing, data processing, and data storage facilities in several countries, which are owned by various parties (each with their own funding sources), and which are collectively operated by the ILT foundation under a joint scientific policy. The efforts of the LSKSP have benefited from funding from the European Research Council, NOVA, NWO, CNRS-INSU, the SURF Co-operative, the UK Science and Technology Funding Council and the Jülich Supercomputing Centre.
\edit1{Construction and installation of VLITE was supported by the Naval Research Laboratory Sustainment Restoration and Maintenance fund.}

%\end{acknowledgements}

The JWST observations used here can be accessed via \dataset{https://doi.org/10.17909/1s0p-qc54}. \edit2{The VLITE data are from VLA project code 18A-338 (P.I.\ R.~Windhorst).}

\software{
SourceExtractor: \citep{Bertin1996} 
\url{https://www.astromatic.net/software/sextractor/} or
\url{https://sextractor.readthedocs.io/en/latest/}}
\software{
PyBDSF \citep{Mohan2015}}
\edit1{\software{
JWST calibration pipeline version 1.7.2 \citep{pipeline} \url{https://zenodo.org/badge/DOI/10.5281/zenodo.7071140.svg}
}}

\facilities{Hubble Space Telescope, James Webb Space Telescope, Mikulski Archive
\url{https://archive.stsci.edu}, MMT/Binospec, MMT/Hectospec, \edit2{Subaru/Hyper-Suprime-Cam,} VLA, LOFAR}

\FloatBarrier
\bibliography{TDFref}{}

\begin{thebibliography}{}
\expandafter\ifx\csname natexlab\endcsname\relax\def\natexlab#1{#1}\fi
\providecommand{\url}[1]{\href{#1}{#1}}
\providecommand{\dodoi}[1]{doi:~\href{http://doi.org/#1}{\nolinkurl{#1}}}
\providecommand{\doeprint}[1]{\href{http://ascl.net/#1}{\nolinkurl{http://ascl.net/#1}}}
\providecommand{\doarXiv}[1]{\href{https://arxiv.org/abs/#1}{\nolinkurl{https://arxiv.org/abs/#1}}}

\bibitem[{{Algera} {et~al.}(2020){Algera}, {van der Vlugt}, {Hodge}, {Smail},
  {Novak}, {Radcliffe}, {Riechers}, {R{\"o}ttgering}, {Smol{\v{c}}i{\'c}}, \&
  {Walter}}]{Algera2020}
{Algera}, H.~S.~B., {van der Vlugt}, D., {Hodge}, J.~A., {et~al.} 2020, \apj,
  903, 139, \dodoi{10.3847/1538-4357/abb77a}

\bibitem[{{Arnouts} {et~al.}(2013){Arnouts}, {Le Floc'h}, {Chevallard},
  {Johnson}, {Ilbert}, {Treyer}, {Aussel}, {Capak}, {Sanders}, {Scoville},
  {McCracken}, {Milliard}, {Pozzetti}, \& {Salvato}}]{Arnouts2013}
{Arnouts}, S., {Le Floc'h}, E., {Chevallard}, J., {et~al.} 2013, \aap, 558,
  A67, \dodoi{10.1051/0004-6361/201321768}

\bibitem[{{Barrufet} {et~al.}(2023){Barrufet}, {Oesch}, {Weibel}, {Brammer},
  {Bezanson}, {Bouwens}, {Fudamoto}, {Gonzalez}, {Gottumukkala}, {Illingworth},
  {Heintz}, {Holden}, {Labbe}, {Magee}, {Naidu}, {Nelson}, {Stefanon}, {Smit},
  {van Dokkum}, {Weaver}, \& {Williams}}]{Barrufet2023}
{Barrufet}, L., {Oesch}, P.~A., {Weibel}, A., {et~al.} 2023, \mnras, 522, 449,
  \dodoi{10.1093/mnras/stad947}

\bibitem[{{Becker} {et~al.}(1994){Becker}, {White}, \& {Helfand}}]{Becker94}
{Becker}, R.~H., {White}, R.~L., \& {Helfand}, D.~J. 1994, in Astronomical
  Society of the Pacific Conference Series, Vol.~61, Astronomical Data Analysis
  Software and Systems III, ed. D.~R. {Crabtree}, R.~J. {Hanisch}, \&
  J.~{Barnes}, 165

\bibitem[{{Bertin} \& {Arnouts}(1996)}]{Bertin1996}
{Bertin}, E., \& {Arnouts}, S. 1996, \aaps, 117, 393,
  \dodoi{10.1051/aas:1996164}

\bibitem[{{Bisigello} {et~al.}(2023){Bisigello}, {Gandolfi}, {Grazian},
  {Rodighiero}, {Costantin}, {Cooray}, {Feltre}, {Gruppioni}, {Hathi},
  {Holwerda}, {Koekemoer}, {Lucas}, {Newman}, {P{\'e}rez-Gonz{\'a}lez}, {Yung},
  {de la Vega}, {Arrabal Haro}, {Bagley}, {Dickinson}, {Finkelstein},
  {Kartaltepe}, {Papovich}, {Pirzkal}, \& {Wilkins}}]{Bisigello2023}
{Bisigello}, L., {Gandolfi}, G., {Grazian}, A., {et~al.} 2023, arXiv e-prints,
  arXiv:2302.12270, \dodoi{10.48550/arXiv.2302.12270}

\bibitem[{{Blundell} \& {Rawlings}(2000)}]{Blundell2000}
{Blundell}, K.~M., \& {Rawlings}, S. 2000, \aj, 119, 1111,
  \dodoi{10.1086/301254}

\bibitem[{{Boquien} {et~al.}(2019){Boquien}, {Burgarella}, {Roehlly}, {Buat},
  {Ciesla}, {Corre}, {Inoue}, \& {Salas}}]{Boquien2019}
{Boquien}, M., {Burgarella}, D., {Roehlly}, Y., {et~al.} 2019, \aap, 622, A103,
  \dodoi{10.1051/0004-6361/201834156}

\bibitem[{{Bruzual} \& {Charlot}(2003)}]{Bruzual2003}
{Bruzual}, G., \& {Charlot}, S. 2003, \mnras, 344, 1000,
  \dodoi{10.1046/j.1365-8711.2003.06897.x}

\bibitem[{Bushouse {et~al.}(2022)Bushouse, Eisenhamer, Dencheva, Davies,
  Greenfield, Morrison, Hodge, Simon, Grumm, Droettboom, Slavich, Sosey, Pauly,
  Miller, Jedrzejewski, Hack, Davis, Crawford, Law, Gordon, Regan, Cara,
  MacDonald, Bradley, Shanahan, Jamieson, Teodoro, \& Williams}]{pipeline}
Bushouse, H., Eisenhamer, J., Dencheva, N., {et~al.} 2022, JWST Calibration
  Pipeline, 1.7.2,  Zenodo, \dodoi{10.5281/zenodo.7071140}

\bibitem[{{Calzetti} {et~al.}(2000){Calzetti}, {Armus}, {Bohlin}, {Kinney},
  {Koornneef}, \& {Storchi-Bergmann}}]{Calzetti2000}
{Calzetti}, D., {Armus}, L., {Bohlin}, R.~C., {et~al.} 2000, \apj, 533, 682,
  \dodoi{10.1086/308692}

\bibitem[{{Chambers} {et~al.}(1996){Chambers}, {Miley}, {van Breugel}, \&
  {Huang}}]{Chambers1996}
{Chambers}, K.~C., {Miley}, G.~K., {van Breugel}, W.~J.~M., \& {Huang}, J.~S.
  1996, \apjs, 106, 215, \dodoi{10.1086/192337}

\bibitem[{{Cheng} {et~al.}(2023){Cheng}, {Huang}, {Smail}, {Yan}, {Cohen},
  {Jansen}, {Windhorst}, {Ma}, {Koekemoer}, {Willmer}, {Willner}, {Diego},
  {Frye}, {Conselice}, {Ferreira}, {Petric}, {Yun}, {Gim}, {Polletta},
  {Duncan}, {Holwerda}, {R{\"o}ttgering}, {Honor}, {Hathi}, {Kamieneski},
  {Adams}, {Coe}, {Broadhurst}, {Summers}, {Tompkins}, {Driver}, {Grogin},
  {Marshall}, {Pirzkal}, {Robotham}, \& {Ryan}}]{Cheng2023}
{Cheng}, C., {Huang}, J.-S., {Smail}, I., {et~al.} 2023, \apjl, 942, L19,
  \dodoi{10.3847/2041-8213/aca9d0}

\bibitem[{{Chilingarian} {et~al.}(2015){Chilingarian}, {Beletsky}, {Moran},
  {Brown}, {McLeod}, \& {Fabricant}}]{Chilingarian2015}
{Chilingarian}, I., {Beletsky}, Y., {Moran}, S., {et~al.} 2015, \pasp, 127,
  406, \dodoi{10.1086/680598}

\bibitem[{{Condon} {et~al.}(1998){Condon}, {Cotton}, {Greisen}, {Yin},
  {Perley}, {Taylor}, \& {Broderick}}]{Condon98}
{Condon}, J.~J., {Cotton}, W.~D., {Greisen}, E.~W., {et~al.} 1998, \aj, 115,
  1693, \dodoi{10.1086/300337}

\bibitem[{{Cotton} {et~al.}(2018){Cotton}, {Condon}, {Kellermann}, {Lacy},
  {Perley}, {Matthews}, {Vernstrom}, {Scott}, \& {Wall}}]{Cotton2018}
{Cotton}, W.~D., {Condon}, J.~J., {Kellermann}, K.~I., {et~al.} 2018, \apj,
  856, 67, \dodoi{10.3847/1538-4357/aaaec4}

\bibitem[{{Draine} {et~al.}(2014){Draine}, {Aniano}, {Krause}, {Groves},
  {Sandstrom}, {Braun}, {Leroy}, {Klaas}, {Linz}, {Rix}, {Schinnerer},
  {Schmiedeke}, \& {Walter}}]{Draine2014}
{Draine}, B.~T., {Aniano}, G., {Krause}, O., {et~al.} 2014, \apj, 780, 172,
  \dodoi{10.1088/0004-637X/780/2/172}

\bibitem[{{Fabricant} {et~al.}(2005){Fabricant}, {Fata}, {Roll}, {Hertz},
  {Caldwell}, {Gauron}, {Geary}, {McLeod}, {Szentgyorgyi}, {Zajac}, {Kurtz},
  {Barberis}, {Bergner}, {Brown}, {Conroy}, {Eng}, {Geller}, {Goddard},
  {Honsa}, {Mueller}, {Mink}, {Ordway}, {Tokarz}, {Woods}, {Wyatt}, {Epps}, \&
  {Dell'Antonio}}]{Fabricant2005}
{Fabricant}, D., {Fata}, R., {Roll}, J., {et~al.} 2005, \pasp, 117, 1411,
  \dodoi{10.1086/497385}

\bibitem[{{Fabricant} {et~al.}(2019){Fabricant}, {Fata}, {Epps}, {Gauron},
  {Mueller}, {Zajac}, {Amato}, {Barberis}, {Bergner}, {Brennan}, {Brown},
  {Chilingarian}, {Geary}, {Kradinov}, {McLeod}, {Smith}, \&
  {Woods}}]{Fabricant2019}
{Fabricant}, D., {Fata}, R., {Epps}, H., {et~al.} 2019, \pasp, 131, 075004,
  \dodoi{10.1088/1538-3873/ab1d78}

\bibitem[{{Fomalont} {et~al.}(2006){Fomalont}, {Kellermann}, {Cowie}, {Capak},
  {Barger}, {Partridge}, {Windhorst}, \& {Richards}}]{Fomalont2006}
{Fomalont}, E.~B., {Kellermann}, K.~I., {Cowie}, L.~L., {et~al.} 2006, \apjs,
  167, 103, \dodoi{10.1086/508169}

\bibitem[{{Gaia Collaboration} {et~al.}(2023){Gaia Collaboration}, {Vallenari},
  {Brown}, {Prusti}, {de Bruijne}, {Arenou}, {Babusiaux}, {Biermann},
  {Creevey}, {Ducourant}, \& et~al.}]{gaia3}
{Gaia Collaboration}, {Vallenari}, A., {Brown}, A.~G.~A., {et~al.} 2023, \aap,
  674, A1, \dodoi{10.1051/0004-6361/202243940}

\bibitem[{{Gilli} {et~al.}(2010){Gilli}, {Vignali}, {Mignoli}, {Iwasawa},
  {Comastri}, \& {Zamorani}}]{Gilli2010}
{Gilli}, R., {Vignali}, C., {Mignoli}, M., {et~al.} 2010, \aap, 519, A92,
  \dodoi{10.1051/0004-6361/201014039}

\bibitem[{{Gregg} {et~al.}(2002){Gregg}, {Lacy}, {White}, {Glikman}, {Helfand},
  {Becker}, \& {Brotherton}}]{Gregg2002}
{Gregg}, M.~D., {Lacy}, M., {White}, R.~L., {et~al.} 2002, \apj, 564, 133,
  \dodoi{10.1086/324145}

\bibitem[{{Haas} {et~al.}(2008){Haas}, {Willner}, {Heymann}, {Ashby}, {Fazio},
  {Wilkes}, {Chini}, \& {Siebenmorgen}}]{Haas2008}
{Haas}, M., {Willner}, S.~P., {Heymann}, F., {et~al.} 2008, \apj, 688, 122,
  \dodoi{10.1086/592085}

\bibitem[{{Herzog} {et~al.}(2016){Herzog}, {Norris}, {Middelberg}, {Seymour},
  {Spitler}, {Emonts}, {Franzen}, {Hunstead}, {Intema}, {Marvil}, {Parker},
  {Sirothia}, {Hurley-Walker}, {Bell}, {Bernardi}, {Bowman}, {Briggs},
  {Cappallo}, {Callingham}, {Deshpande}, {Dwarakanath}, {For}, {Greenhill},
  {Hancock}, {Hazelton}, {Hindson}, {Johnston-Hollitt}, {Kapi{\'n}ska},
  {Kaplan}, {Lenc}, {Lonsdale}, {McKinley}, {McWhirter}, {Mitchell}, {Morales},
  {Morgan}, {Morgan}, {Oberoi}, {Offringa}, {Ord}, {Prabu}, {Procopio}, {Udaya
  Shankar}, {Srivani}, {Staveley-Smith}, {Subrahmanyan}, {Tingay}, {Wayth},
  {Webster}, {Williams}, {Williams}, {Wu}, {Zheng}, {Bannister}, {Chippendale},
  {Harvey-Smith}, {Heywood}, {Indermuehle}, {Popping}, {Sault}, \&
  {Whiting}}]{Herzog2016}
{Herzog}, A., {Norris}, R.~P., {Middelberg}, E., {et~al.} 2016, \aap, 593,
  A130, \dodoi{10.1051/0004-6361/201527000}

\bibitem[{{Heywood} {et~al.}(2021){Heywood}, {Murphy}, {Jim{\'e}nez-Andrade},
  {Armus}, {Cotton}, {DeCoursey}, {Dickinson}, {Lazio}, {Momjian}, {Penner},
  {Smail}, \& {Smirnov}}]{Heywood2021}
{Heywood}, I., {Murphy}, E.~J., {Jim{\'e}nez-Andrade}, E.~F., {et~al.} 2021,
  \apj, 910, 105, \dodoi{10.3847/1538-4357/abdf61}

\bibitem[{{Hyun} {et~al.}(2023){Hyun}, {Im}, {Smail}, {Cotton}, {Birkin},
  {Kikuta}, {Shim}, {Willmer}, {Condon}, {Windhorst}, {Cohen}, {Jansen}, {Ly},
  {Matsuda}, {Fazio}, {Swinbank}, \& {Yan}}]{Hyun2023}
{Hyun}, M., {Im}, M., {Smail}, I.~R., {et~al.} 2023, \apjs, 264, 19,
  \dodoi{10.3847/1538-4365/ac9bf4}

\bibitem[{{Ivison} {et~al.}(2007){Ivison}, {Chapman}, {Faber}, {Smail},
  {Biggs}, {Conselice}, {Wilson}, {Salim}, {Huang}, \& {Willner}}]{Ivison2007}
{Ivison}, R.~J., {Chapman}, S.~C., {Faber}, S.~M., {et~al.} 2007, \apjl, 660,
  L77, \dodoi{10.1086/517917}

\bibitem[{{Jansen} \& {Windhorst}(2018)}]{Jansen2018}
{Jansen}, R.~A., \& {Windhorst}, R.~A. 2018, \pasp, 130, 124001,
  \dodoi{10.1088/1538-3873/aae476}

\bibitem[{{Jones} {et~al.}(2017){Jones}, {K{\"o}hler}, {Ysard}, {Bocchio}, \&
  {Verstraete}}]{Jones2017}
{Jones}, A.~P., {K{\"o}hler}, M., {Ysard}, N., {Bocchio}, M., \& {Verstraete},
  L. 2017, \aap, 602, A46, \dodoi{10.1051/0004-6361/201630225}

\bibitem[{{Komatsu} {et~al.}(2011){Komatsu}, {Smith}, {Dunkley}, {Bennett},
  {Gold}, {Hinshaw}, {Jarosik}, {Larson}, {Nolta}, {Page}, {Spergel},
  {Halpern}, {Hill}, {Kogut}, {Limon}, {Meyer}, {Odegard}, {Tucker}, {Weiland},
  {Wollack}, \& {Wright}}]{Komatsu2011}
{Komatsu}, E., {Smith}, K.~M., {Dunkley}, J., {et~al.} 2011, \apjs, 192, 18,
  \dodoi{10.1088/0067-0049/192/2/18}

\bibitem[{{Kondapally} {et~al.}(2021){Kondapally}, {Best}, {Hardcastle},
  {Nisbet}, {Bonato}, {Sabater}, {Duncan}, {McCheyne}, {Cochrane}, {Bowler},
  {Williams}, {Shimwell}, {Tasse}, {Croston}, {Goyal}, {Jamrozy}, {Jarvis},
  {Mahatma}, {R{\"o}ttgering}, {Smith}, {Wo{\l}owska}, {Bondi}, {Brienza},
  {Brown}, {Br{\"u}ggen}, {Chambers}, {Garrett}, {G{\"u}rkan}, {Huber},
  {Kunert-Bajraszewska}, {Magnier}, {Mingo}, {Mostert},
  {Nikiel-Wroczy{\'n}ski}, {O'Sullivan}, {Paladino}, {Ploeckinger}, {Prandoni},
  {Rosenthal}, {Schwarz}, {Shulevski}, {Wagenveld}, \& {Wang}}]{Kondapally2021}
{Kondapally}, R., {Best}, P.~N., {Hardcastle}, M.~J., {et~al.} 2021, \aap, 648,
  A3, \dodoi{10.1051/0004-6361/202038813}

\bibitem[{{Madau} \& {Dickinson}(2014)}]{Madau2014}
{Madau}, P., \& {Dickinson}, M. 2014, \araa, 52, 415,
  \dodoi{10.1146/annurev-astro-081811-125615}

\bibitem[{{Magnelli} {et~al.}(2023){Magnelli}, {G{\'o}mez-Guijarro}, {Elbaz},
  {Daddi}, {Papovich}, {Shen}, {Arrabal Haro}, {Bagley}, {Bell}, {Buat},
  {Costantin}, {Dickinson}, {Finkelstein}, {Gardner}, {Jim{\'e}nez-Andrade},
  {Kartaltepe}, {Koekemoer}, {Lyu}, {P{\'e}rez-Gonz{\'a}lez}, {Pirzkal},
  {Tacchella}, {de la Vega}, {Wuyts}, {Yang}, {Yung}, \&
  {Zavala}}]{Magnelli2023}
{Magnelli}, B., {G{\'o}mez-Guijarro}, C., {Elbaz}, D., {et~al.} 2023, arXiv
  e-prints, arXiv:2305.19331, \dodoi{10.48550/arXiv.2305.19331}

\bibitem[{{Mahajan} {et~al.}(2019){Mahajan}, {Ashby}, {Willner}, {Barmby},
  {Fazio}, {Maragkoudakis}, {Raychaudhury}, \& {Zezas}}]{Mahajan2019}
{Mahajan}, S., {Ashby}, M.~L.~N., {Willner}, S.~P., {et~al.} 2019, \mnras, 482,
  560, \dodoi{10.1093/mnras/sty2699}

\bibitem[{{McLeod} {et~al.}(2012){McLeod}, {Fabricant}, {Nystrom}, {McCracken},
  {Amato}, {Bergner}, {Brown}, {Burke}, {Chilingarian}, {Conroy}, {Curley},
  {Furesz}, {Geary}, {Hertz}, {Holwell}, {Matthews}, {Norton}, {Park}, {Roll},
  {Zajac}, {Epps}, \& {Martini}}]{McLeod2012}
{McLeod}, B., {Fabricant}, D., {Nystrom}, G., {et~al.} 2012, \pasp, 124, 1318,
  \dodoi{10.1086/669044}

\bibitem[{{Mignoli} {et~al.}(2013){Mignoli}, {Vignali}, {Gilli}, {Comastri},
  {Zamorani}, {Bolzonella}, {Bongiorno}, {Lamareille}, {Nair}, {Pozzetti},
  {Lilly}, {Carollo}, {Contini}, {Kneib}, {Le F{\`e}vre}, {Mainieri},
  {Renzini}, {Scodeggio}, {Bardelli}, {Caputi}, {Cucciati}, {de la Torre}, {de
  Ravel}, {Franzetti}, {Garilli}, {Iovino}, {Kampczyk}, {Knobel},
  {Kova{\v{c}}}, {Le Borgne}, {Le Brun}, {Maier}, {Pell{\`o}}, {Peng}, {Perez
  Montero}, {Presotto}, {Silverman}, {Tanaka}, {Tasca}, {Tresse}, {Vergani},
  {Zucca}, {Bordoloi}, {Cappi}, {Cimatti}, {Koekemoer}, {McCracken}, {Moresco},
  \& {Welikala}}]{Mignoli2013}
{Mignoli}, M., {Vignali}, C., {Gilli}, R., {et~al.} 2013, \aap, 556, A29,
  \dodoi{10.1051/0004-6361/201220846}

\bibitem[{{Miley} \& {De Breuck}(2008)}]{Miley2008}
{Miley}, G., \& {De Breuck}, C. 2008, \aapr, 15, 67,
  \dodoi{10.1007/s00159-007-0008-z}

\bibitem[{{Mohan} \& {Rafferty}(2015)}]{Mohan2015}
{Mohan}, N., \& {Rafferty}, D. 2015, {PyBDSF: Python Blob Detection and Source
  Finder}, Astrophysics Source Code Library, record ascl:1502.007.
\newblock \doeprint{1502.007}

\bibitem[{{Murphy} {et~al.}(2011){Murphy}, {Condon}, {Schinnerer}, {Kennicutt},
  {Calzetti}, {Armus}, {Helou}, {Turner}, {Aniano}, {Beir{\~a}o}, {Bolatto},
  {Brandl}, {Croxall}, {Dale}, {Donovan Meyer}, {Draine}, {Engelbracht},
  {Hunt}, {Hao}, {Koda}, {Roussel}, {Skibba}, \& {Smith}}]{Murphy2011}
{Murphy}, E.~J., {Condon}, J.~J., {Schinnerer}, E., {et~al.} 2011, \apj, 737,
  67, \dodoi{10.1088/0004-637X/737/2/67}

\bibitem[{{Newman} {et~al.}(2013){Newman}, {Cooper}, {Davis}, {Faber}, {Coil},
  {Guhathakurta}, {Koo}, {Phillips}, {Conroy}, {Dutton}, {Finkbeiner}, {Gerke},
  {Rosario}, {Weiner}, {Willmer}, {Yan}, {Harker}, {Kassin}, {Konidaris},
  {Lai}, {Madgwick}, {Noeske}, {Wirth}, {Connolly}, {Kaiser}, {Kirby},
  {Lemaux}, {Lin}, {Lotz}, {Luppino}, {Marinoni}, {Matthews}, {Metevier}, \&
  {Schiavon}}]{Newman2013}
{Newman}, J.~A., {Cooper}, M.~C., {Davis}, M., {et~al.} 2013, \apjs, 208, 5,
  \dodoi{10.1088/0067-0049/208/1/5}

\bibitem[{{Oort}(1988)}]{Oort1988}
{Oort}, M.~J.~A. 1988, \aap, 193, 5

\bibitem[{{Owen}(2018)}]{Owen2018}
{Owen}, F.~N. 2018, \apjs, 235, 34, \dodoi{10.3847/1538-4365/aab4a1}

\bibitem[{{P{\'e}rez-Gonz{\'a}lez} {et~al.}(2023){P{\'e}rez-Gonz{\'a}lez},
  {Barro}, {Annunziatella}, {Costantin}, {Garc{\'\i}a-Argum{\'a}nez},
  {McGrath}, {M{\'e}rida}, {Zavala}, {Haro}, {Bagley}, {Backhaus}, {Behroozi},
  {Bell}, {Bisigello}, {Buat}, {Calabr{\`o}}, {Casey}, {Cleri}, {Coogan},
  {Cooper}, {Cooray}, {Dekel}, {Dickinson}, {Elbaz}, {Ferguson}, {Finkelstein},
  {Fontana}, {Franco}, {Gardner}, {Giavalisco}, {G{\'o}mez-Guijarro},
  {Grazian}, {Grogin}, {Guo}, {Huertas-Company}, {Jogee}, {Kartaltepe},
  {Kewley}, {Kirkpatrick}, {Kocevski}, {Koekemoer}, {Long}, {Lotz}, {Lucas},
  {Papovich}, {Pirzkal}, {Ravindranath}, {Somerville}, {Tacchella}, {Trump},
  {Wang}, {Wilkins}, {Wuyts}, {Yang}, \& {Yung}}]{Perez-Gonzalez2023}
{P{\'e}rez-Gonz{\'a}lez}, P.~G., {Barro}, G., {Annunziatella}, M., {et~al.}
  2023, \apjl, 946, L16, \dodoi{10.3847/2041-8213/acb3a5}

\bibitem[{{Peters} {et~al.}(2021){Peters}, {Polisensky}, {Brisken}, {Cotton},
  {Clarke}, {Giacintucci}, \& {Kassim}}]{vlite}
{Peters}, W., {Polisensky}, E., {Brisken}, W., {et~al.} 2021, in American
  Astronomical Society Meeting Abstracts, Vol.~53, American Astronomical
  Society Meeting Abstracts, 211.06

\bibitem[{{Polletta} {et~al.}(2007){Polletta}, {Tajer}, {Maraschi},
  {Trinchieri}, {Lonsdale}, {Chiappetti}, {Andreon}, {Pierre}, {Le F{\`e}vre},
  {Zamorani}, {Maccagni}, {Garcet}, {Surdej}, {Franceschini}, {Alloin},
  {Shupe}, {Surace}, {Fang}, {Rowan-Robinson}, {Smith}, \&
  {Tresse}}]{Polletta2007}
{Polletta}, M., {Tajer}, M., {Maraschi}, L., {et~al.} 2007, \apj, 663, 81,
  \dodoi{10.1086/518113}

\bibitem[{{Renzini} \& {Peng}(2015)}]{Renzini2015}
{Renzini}, A., \& {Peng}, Y.-j. 2015, \apjl, 801, L29,
  \dodoi{10.1088/2041-8205/801/2/L29}

\bibitem[{{Rigby} {et~al.}(2023){Rigby}, {Perrin}, {McElwain}, {Kimble},
  {Friedman}, {Lallo}, {Doyon}, {Feinberg}, {Ferruit}, {Glasse}, \&
  et~al.}]{Rigby2023}
{Rigby}, J., {Perrin}, M., {McElwain}, M., {et~al.} 2023, \pasp, 135, 048001,
  \dodoi{10.1088/1538-3873/acb293}

\bibitem[{{Rodighiero} {et~al.}(2011){Rodighiero}, {Daddi}, {Baronchelli},
  {Cimatti}, {Renzini}, {Aussel}, {Popesso}, {Lutz}, {Andreani}, {Berta},
  {Cava}, {Elbaz}, {Feltre}, {Fontana}, {F{\"o}rster Schreiber},
  {Franceschini}, {Genzel}, {Grazian}, {Gruppioni}, {Ilbert}, {Le Floch},
  {Magdis}, {Magliocchetti}, {Magnelli}, {Maiolino}, {McCracken}, {Nordon},
  {Poglitsch}, {Santini}, {Pozzi}, {Riguccini}, {Tacconi}, {Wuyts}, \&
  {Zamorani}}]{Rodighiero2011}
{Rodighiero}, G., {Daddi}, E., {Baronchelli}, I., {et~al.} 2011, \apjl, 739,
  L40, \dodoi{10.1088/2041-8205/739/2/L40}

\bibitem[{{Russell} {et~al.}(2008){Russell}, {Ryan}, {Cohen}, {Windhorst}, \&
  {Waddington}}]{Russell2008}
{Russell}, J., {Ryan}, R.~E., J., {Cohen}, S.~H., {Windhorst}, R.~A., \&
  {Waddington}, I. 2008, \apjs, 179, 306, \dodoi{10.1086/592045}

\bibitem[{{Sabater} {et~al.}(2019){Sabater}, {Best}, {Hardcastle}, {Shimwell},
  {Tasse}, {Williams}, {Br{\"u}ggen}, {Cochrane}, {Croston}, {de Gasperin},
  {Duncan}, {G{\"u}rkan}, {Mechev}, {Morabito}, {Prandoni}, {R{\"o}ttgering},
  {Smith}, {Harwood}, {Mingo}, {Mooney}, \& {Saxena}}]{Sabater2019}
{Sabater}, J., {Best}, P.~N., {Hardcastle}, M.~J., {et~al.} 2019, \aap, 622,
  A17, \dodoi{10.1051/0004-6361/201833883}

\bibitem[{{Salpeter}(1955)}]{Salpeter1955}
{Salpeter}, E.~E. 1955, \apj, 121, 161, \dodoi{10.1086/145971}

\bibitem[{{Sanders} {et~al.}(1989){Sanders}, {Phinney}, {Neugebauer}, {Soifer},
  \& {Matthews}}]{Sanders1989}
{Sanders}, D.~B., {Phinney}, E.~S., {Neugebauer}, G., {Soifer}, B.~T., \&
  {Matthews}, K. 1989, \apj, 347, 29, \dodoi{10.1086/168094}

\bibitem[{{Sawicki}(2002)}]{Sawicki02}
{Sawicki}, M. 2002, \aj, 124, 3050, \dodoi{10.1086/344682}

\bibitem[{{Saxena} {et~al.}(2019){Saxena}, {R{\"o}ttgering}, {Duncan}, {Hill},
  {Best}, {Indahl}, {Marinello}, {Overzier}, {Pentericci}, {Prandoni},
  {Dannerbauer}, \& {Barrena}}]{Saxena2019}
{Saxena}, A., {R{\"o}ttgering}, H.~J.~A., {Duncan}, K.~J., {et~al.} 2019,
  \mnras, 489, 5053, \dodoi{10.1093/mnras/stz2516}

\bibitem[{{Seymour} {et~al.}(2007){Seymour}, {Stern}, {De Breuck}, {Vernet},
  {Rettura}, {Dickinson}, {Dey}, {Eisenhardt}, {Fosbury}, {Lacy}, {McCarthy},
  {Miley}, {Rocca-Volmerange}, {R{\"o}ttgering}, {Stanford}, {Teplitz}, {van
  Breugel}, \& {Zirm}}]{Seymour2007}
{Seymour}, N., {Stern}, D., {De Breuck}, C., {et~al.} 2007, \apjs, 171, 353,
  \dodoi{10.1086/517887}

\bibitem[{{Shimwell} {et~al.}(2022){Shimwell}, {Hardcastle}, {Tasse}, {Best},
  {R{\"o}ttgering}, {Williams}, {Botteon}, {Drabent}, {Mechev}, {Shulevski},
  {van Weeren}, {Bester}, {Br{\"u}ggen}, {Brunetti}, {Callingham}, {Chy{\.z}y},
  {Conway}, {Dijkema}, {Duncan}, {de Gasperin}, {Hale}, {Haverkorn}, {Hugo},
  {Jackson}, {Mevius}, {Miley}, {Morabito}, {Morganti}, {Offringa}, {Oonk},
  {Rafferty}, {Sabater}, {Smith}, {Schwarz}, {Smirnov}, {O'Sullivan},
  {Vedantham}, {White}, {Albert}, {Alegre}, {Asabere}, {Bacon}, {Bonafede},
  {Bonnassieux}, {Brienza}, {Bilicki}, {Bonato}, {Calistro Rivera}, {Cassano},
  {Cochrane}, {Croston}, {Cuciti}, {Dallacasa}, {Danezi}, {Dettmar}, {Di
  Gennaro}, {Edler}, {En{\ss}lin}, {Emig}, {Franzen}, {Garc{\'\i}a-Vergara},
  {Grange}, {G{\"u}rkan}, {Hajduk}, {Heald}, {Heesen}, {Hoang}, {Hoeft},
  {Horellou}, {Iacobelli}, {Jamrozy}, {Jeli{\'c}}, {Kondapally}, {Kukreti},
  {Kunert-Bajraszewska}, {Magliocchetti}, {Mahatma}, {Ma{\l}ek}, {Mandal},
  {Massaro}, {Meyer-Zhao}, {Mingo}, {Mostert}, {Nair}, {Nakoneczny},
  {Nikiel-Wroczy{\'n}ski}, {Orr{\'u}}, {Pajdosz-{\'S}mierciak}, {Pasini},
  {Prandoni}, {van Piggelen}, {Rajpurohit}, {Retana-Montenegro}, {Riseley},
  {Rowlinson}, {Saxena}, {Schrijvers}, {Sweijen}, {Siewert}, {Timmerman},
  {Vaccari}, {Vink}, {West}, {Wo{\l}owska}, {Zhang}, \& {Zheng}}]{Shimwell2022}
{Shimwell}, T.~W., {Hardcastle}, M.~J., {Tasse}, C., {et~al.} 2022, \aap, 659,
  A1, \dodoi{10.1051/0004-6361/202142484}

\bibitem[{{Smol{\v{c}}i{\'c}} {et~al.}(2017{\natexlab{a}}){Smol{\v{c}}i{\'c}},
  {Novak}, {Bondi}, {Ciliegi}, {Mooley}, {Schinnerer}, {Zamorani}, {Navarrete},
  {Bourke}, {Karim}, {Vardoulaki}, {Leslie}, {Delhaize}, {Carilli}, {Myers},
  {Baran}, {Delvecchio}, {Miettinen}, {Banfield}, {Balokovi{\'c}}, {Bertoldi},
  {Capak}, {Frail}, {Hallinan}, {Hao}, {Herrera Ruiz}, {Horesh}, {Ilbert},
  {Intema}, {Jeli{\'c}}, {Kl{\"o}ckner}, {Krpan}, {Kulkarni}, {McCracken},
  {Laigle}, {Middleberg}, {Murphy}, {Sargent}, {Scoville}, \&
  {Sheth}}]{Smolcic2017data}
{Smol{\v{c}}i{\'c}}, V., {Novak}, M., {Bondi}, M., {et~al.} 2017{\natexlab{a}},
  \aap, 602, A1, \dodoi{10.1051/0004-6361/201628704}

\bibitem[{{Smol{\v{c}}i{\'c}} {et~al.}(2017{\natexlab{b}}){Smol{\v{c}}i{\'c}},
  {Delvecchio}, {Zamorani}, {Baran}, {Novak}, {Delhaize}, {Schinnerer},
  {Berta}, {Bondi}, {Ciliegi}, {Capak}, {Civano}, {Karim}, {Le Fevre},
  {Ilbert}, {Laigle}, {Marchesi}, {McCracken}, {Tasca}, {Salvato}, \&
  {Vardoulaki}}]{Smolcic2017id}
{Smol{\v{c}}i{\'c}}, V., {Delvecchio}, I., {Zamorani}, G., {et~al.}
  2017{\natexlab{b}}, \aap, 602, A2, \dodoi{10.1051/0004-6361/201630223}

\bibitem[{{Songaila} {et~al.}(2018){Songaila}, {Hu}, {Barger}, {Cowie},
  {Hasinger}, {Rosenwasser}, \& {Waters}}]{Songaila2018}
{Songaila}, A., {Hu}, E.~M., {Barger}, A.~J., {et~al.} 2018, \apj, 859, 91,
  \dodoi{10.3847/1538-4357/aac021}

\bibitem[{{Speagle} {et~al.}(2014){Speagle}, {Steinhardt}, {Capak}, \&
  {Silverman}}]{Speagle2014}
{Speagle}, J.~S., {Steinhardt}, C.~L., {Capak}, P.~L., \& {Silverman}, J.~D.
  2014, \apjs, 214, 15, \dodoi{10.1088/0067-0049/214/2/15}

\bibitem[{{Spinrad} {et~al.}(1985){Spinrad}, {Djorgovski}, {Marr}, \&
  {Aguilar}}]{Spinrad1985}
{Spinrad}, H., {Djorgovski}, S., {Marr}, J., \& {Aguilar}, L. 1985, \pasp, 97,
  932, \dodoi{10.1086/131647}

\bibitem[{{Stalevski} {et~al.}(2012){Stalevski}, {Fritz}, {Baes}, {Nakos}, \&
  {Popovi{\'c}}}]{Stalevski2012}
{Stalevski}, M., {Fritz}, J., {Baes}, M., {Nakos}, T., \& {Popovi{\'c}},
  L.~{\v{C}}. 2012, \mnras, 420, 2756, \dodoi{10.1111/j.1365-2966.2011.19775.x}

\bibitem[{{Stalevski} {et~al.}(2016){Stalevski}, {Ricci}, {Ueda}, {Lira},
  {Fritz}, \& {Baes}}]{Stalevski2016}
{Stalevski}, M., {Ricci}, C., {Ueda}, Y., {et~al.} 2016, \mnras, 458, 2288,
  \dodoi{10.1093/mnras/stw444}

\bibitem[{{Strazzullo} {et~al.}(2010){Strazzullo}, {Pannella}, {Owen},
  {Bender}, {Morrison}, {Wang}, \& {Shupe}}]{Strazzullo2010}
{Strazzullo}, V., {Pannella}, M., {Owen}, F.~N., {et~al.} 2010, \apj, 714,
  1305, \dodoi{10.1088/0004-637X/714/2/1305}

\bibitem[{{Taylor} {et~al.}(2023){Taylor}, {Barger}, {Cowie}, {Hasinger}, {Hu},
  \& {Songaila}}]{Taylor2023}
{Taylor}, A.~J., {Barger}, A.~J., {Cowie}, L.~L., {et~al.} 2023, \apjs, 266,
  24, \dodoi{10.3847/1538-4365/accd70}

\bibitem[{{Tompkins} {et~al.}(2023){Tompkins}, {Driver}, {Robotham},
  {Windhorst}, {Lagos}, {Vernstrom}, \& {Hopkins}}]{Tompkins2023}
{Tompkins}, S.~A., {Driver}, S.~P., {Robotham}, A. S.~G., {et~al.} 2023,
  \mnras, 521, 332, \dodoi{10.1093/mnras/stad116}

\bibitem[{{van Haarlem} {et~al.}(2013){van Haarlem}, {Wise}, {Gunst}, {Heald},
  {McKean}, {Hessels}, {de Bruyn}, {Nijboer}, {Swinbank}, {Fallows}, \&
  et~al.}]{lofar}
{van Haarlem}, M.~P., {Wise}, M.~W., {Gunst}, A.~W., {et~al.} 2013, \aap, 556,
  A2, \dodoi{10.1051/0004-6361/201220873}

\bibitem[{{Williams} {et~al.}(2009){Williams}, {Quadri}, {Franx}, {van Dokkum},
  \& {Labb{\'e}}}]{Williams2009}
{Williams}, R.~J., {Quadri}, R.~F., {Franx}, M., {van Dokkum}, P., \&
  {Labb{\'e}}, I. 2009, \apj, 691, 1879, \dodoi{10.1088/0004-637X/691/2/1879}

\bibitem[{{Willmer} {et~al.}(2023){Willmer}, {Chun Ly}, {Satoshi Kikuta}, {S.
  A. Kattner}, {Rolf A. Jansen}, {Seth H. Cohen}, {Rogier A. Windhorst}, {Ian
  Smail}, {Scott Tompkins}, {John F. Beacom}, {Cheng Cheng}, {Christopher J.
  Conselice}, {Brenda L. Frye}, {Anton M. Koekemoer}, {Nimish Hathi}, {Minhee
  Hyun}, {Myungshin Im}, {S.\ P.\ Willner}, {X. Zhao}, {Walter A. Brisken}, {F.
  Civano}, {William Cotton}, {G\"unther Hasinger}, {W. Peter Maksym}, \&
  {Marcia J. Rieke}}]{Willmer2023}
{Willmer}, C.~N.~A., {Chun Ly}, {Satoshi Kikuta}, {et~al.} 2023, \apj,
  submitted

\bibitem[{{Willner} {et~al.}(2012){Willner}, {Ashby}, {Barmby}, {Chapman},
  {Coil}, {Cooper}, {Huang}, {Ivison}, \& {Koo}}]{Willner2012}
{Willner}, S.~P., {Ashby}, M.~L.~N., {Barmby}, P., {et~al.} 2012, \apj, 756,
  72, \dodoi{10.1088/0004-637X/756/1/72}

\bibitem[{{Windhorst} {et~al.}(1990){Windhorst}, {Mathis}, \&
  {Neuschaefer}}]{Windhorst1990}
{Windhorst}, R., {Mathis}, D., \& {Neuschaefer}, L. 1990, in Astronomical
  Society of the Pacific Conference Series, Vol.~10, Evolution of the Universe
  of Galaxies, ed. R.~G. {Kron}, 389--403

\bibitem[{{Windhorst}(2003)}]{Windhorst2003}
{Windhorst}, R.~A. 2003, \nar, 47, 357, \dodoi{10.1016/S1387-6473(03)00045-9}

\bibitem[{{Windhorst} {et~al.}(1995){Windhorst}, {Fomalont}, {Kellermann},
  {Partridge}, {Richards}, {Franklin}, {Pascarelle}, \&
  {Griffiths}}]{Windhorst1995}
{Windhorst}, R.~A., {Fomalont}, E.~B., {Kellermann}, K.~I., {et~al.} 1995,
  \nat, 375, 471, \dodoi{10.1038/375471a0}

\bibitem[{{Windhorst} {et~al.}(1985){Windhorst}, {Miley}, {Owen}, {Kron}, \&
  {Koo}}]{Windhorst1985}
{Windhorst}, R.~A., {Miley}, G.~K., {Owen}, F.~N., {Kron}, R.~G., \& {Koo},
  D.~C. 1985, \apj, 289, 494, \dodoi{10.1086/162911}

\bibitem[{{Windhorst} {et~al.}(1984){Windhorst}, {van Heerde}, \&
  {Katgert}}]{Windhorst1984}
{Windhorst}, R.~A., {van Heerde}, G.~M., \& {Katgert}, P. 1984, \aaps, 58, 1

\bibitem[{{Windhorst} {et~al.}(2023){Windhorst}, {Cohen}, {Jansen}, {Summers},
  {Tompkins}, {Conselice}, {Driver}, {Yan}, {Coe}, {Frye}, {Grogin},
  {Koekemoer}, {Marshall}, {O'Brien}, {Pirzkal}, {Robotham}, {Ryan}, {Willmer},
  {Carleton}, {Diego}, {Keel}, {Porto}, {Redshaw}, {Scheller}, {Wilkins},
  {Willner}, {Zitrin}, {Adams}, {Austin}, {Arendt}, {Beacom}, {Bhatawdekar},
  {Bradley}, {Broadhurst}, {Cheng}, {Civano}, {Dai}, {Dole}, {D'Silva},
  {Duncan}, {Fazio}, {Ferrami}, {Ferreira}, {Finkelstein}, {Furtak}, {Gim},
  {Griffiths}, {Hammel}, {Harrington}, {Hathi}, {Holwerda}, {Honor}, {Huang},
  {Hyun}, {Im}, {Joshi}, {Kamieneski}, {Kelly}, {Larson}, {Li}, {Lim}, {Ma},
  {Maksym}, {Manzoni}, {Meena}, {Milam}, {Nonino}, {Pascale}, {Petric},
  {Pierel}, {del Carmen Polletta}, {R{\"o}ttgering}, {Rutkowski}, {Smail},
  {Straughn}, {Strolger}, {Swirbul}, {Trussler}, {Wang}, {Welch}, {B. Wyithe},
  {Yun}, {Zackrisson}, {Zhang}, \& {Zhao}}]{Windhorst2023}
{Windhorst}, R.~A., {Cohen}, S.~H., {Jansen}, R.~A., {et~al.} 2023, \aj, 165,
  13, \dodoi{10.3847/1538-3881/aca163}

\bibitem[{{Yang} {et~al.}(2020){Yang}, {Boquien}, {Buat}, {Burgarella},
  {Ciesla}, {Duras}, {Stalevski}, {Brandt}, \& {Papovich}}]{Yang2020}
{Yang}, G., {Boquien}, M., {Buat}, V., {et~al.} 2020, \mnras, 491, 740,
  \dodoi{10.1093/mnras/stz3001}

\bibitem[{{Yang} {et~al.}(2022){Yang}, {Boquien}, {Brandt}, {Buat},
  {Burgarella}, {Ciesla}, {Lehmer}, {Ma{\l}ek}, {Mountrichas}, {Papovich},
  {Pons}, {Stalevski}, {Theul{\'e}}, \& {Zhu}}]{Yang2022}
{Yang}, G., {Boquien}, M., {Brandt}, W.~N., {et~al.} 2022, \apj, 927, 192,
  \dodoi{10.3847/1538-4357/ac4971}

\bibitem[{{Zinn} {et~al.}(2011){Zinn}, {Middelberg}, \& {Ibar}}]{Zinn2011}
{Zinn}, P.~C., {Middelberg}, E., \& {Ibar}, E. 2011, \aap, 531, A14,
  \dodoi{10.1051/0004-6361/201016264}

\end{thebibliography}
\bibliographystyle{aasjournal}

%\clearpage
\appendix
\restartappendixnumbering

\section{Estimating the Size-Dependent Incompleteness of the
  \hbox{3~GHz} Sample}
\label{sa:sb}

The objective of this paper is to examine JWST/NIRCam observations of radio sources in a typical sensitive, high-angular-resolution survey.  While incompleteness of the radio survey itself is not directly relevant to this paper's conclusions, it is nevertheless worth quantifying what that incompleteness might be. Interferometric radio surveys are limited by {\em both} point-source
{\em and} surface-brightness (SB) sensitivities.  The longest
baselines are sensitive to point sources, but sufficient shorter baselines are needed
for good SB-sensitivity (\eg, Fig.~7 of \citealt{Windhorst1984} and references
therein.)

The survey used here \citep{Hyun2023}
was based on a combination of VLA A and B arrays. The 44 hours of
A-array data achieved rms noise of $\sim$1\,\mmJy\,beam$^{-1}$. The beam was nearly
circular with $\rm FWHM = 0\farcs7$.  This high resolution
provided the position accuracy required for reliable
identifications on the high-resolution NIRCam images.  The 4 hours of B-array data achieved rms
noise of 3.2\,\mmJy\,beam$^{-1}$ in a beam of 2\farcs3$\times$2\farcs1. While this improved the SB
sensitivity, the combination still lacks sensitivity in the shorter baselines, especially those
that the C-array would have provided. As a consequence, the
\citeauthor{Hyun2023} source list is necessarily missing some faint,
low-SB radio sources.

There are four ways to estimate the source incompleteness, two empirical and two theoretical.
The simplest empirical estimate comes from the \citet{Cotton2018} 3\,GHz survey of the Lockman Hole.  That survey was about the
same depth as the \citet{Hyun2023} survey used
here, but it included C-array data and a higher proportion of B-array data than the \citet{Hyun2023} survey. Adding the C-array data gave 10\% more sources than the A+B data alone \citep{Cotton2018}.  This is a lower limit to the SB incompleteness because of the  greater proportion of B-array data in the \citeauthor{Cotton2018} survey.

A second empirical incompleteness estimate comes from convolving the \citet{Hyun2023} 3\,GHz image with a kernel to obtain the ``B-array-only'' beam of
2\farcs3$\times$2\farcs1 FWHM with a resulting rms image noise of
1.5\,\mmJy\,beam$^{-1}$. (The lower noise than the pure B-array image comes from including short baselines contributed by the A-array.) Above its 5$\sigma$ level, there are 12 low-SB sources
inside the NIRCam footprint that are not in the  original \citet{Hyun2023} radio
catalog. This adds 19\% to our list of 63 sources.  However, four sources look to be sidelobes and have no NIRCam counterparts, so the added fraction is probably closer to 13\%.  (Of the eight sources that do not look like sidelobes, seven have obvious spiral- or irregular-galaxy NIRCam counterparts, and one has a possible but doubtful NIRCam counterpart.)  Given the small numbers, adding eight sources to the original 63 is consistent with the 10\% estimate based on the \citet{Cotton2018} survey.

Theoretical estimates of incompleteness depend on knowing the true angular-size distribution of the radio population.  In essence, at a given flux-density limit, a radio survey is complete for sizes smaller than some value. If the completeness limit is known, the angular-size distribution reveals how many sources with sizes larger than that limit must exist.  The angular-size distribution of radio sources is dependent on flux density, but below 1\,mJy, sources larger than 5\arcsec\ are rare \citep[][their Fig.~2]{Windhorst1990}. 
Based on the 1.4\,GHz flux-dependent angular size distribution work
by \citet{Oort1988}---which was measured with VLA A-array for
a sample defined at 3\,km baselines and thereby constructed to include sources up to
30\arcsec\ in angular size---\citet{Windhorst1990} showed that the fraction of radio sources with size larger than some value $\Psi$ is
\begin{equation} 
h(\Psi) = \exp\ [-(\ln 2) (\Psi/\langle\Psi\rangle)^{0.62}] \quad. 
\label{eq:h} 
\end{equation} 
Here, $\langle\Psi\rangle$ is the
median angular size of the population, which is a function of flux density.\footnote{The median angular size at different flux densities has been well determined over time with radio interferometers at
increasing resolution and sensitivity \citep[\eg, Fig.~4
of][]{Windhorst2003}.  Between 30 \mmJy\ and 1 Jy, the relation is well represented
by a power-law:
\begin{equation} 
\langle\Psi\rangle\ = 2\farcs0 \,S({\rm 1.4\,GHz})^{0.30} \quad,
\label{eq:eq1} 
\end{equation} 
where S({\rm 1.4\,GHz}) is measured in mJy. The median
high-frequency spectral index for a sample fainter than 30\,\mmJy\ is
$\alpha\simeq-0.4$ \citep[][their Fig.~1a]{Tompkins2023}, which gives a multiplier of 1.356 to convert 3\,GHz flux densities to 1.4\,GHz.  The predicted median angular size above our flux limit is then $\langle\Psi\rangle = 0\farcs45$, reasonably close to the observed 0\farcs30 given that the equation was derived at much higher flux densities.}  (By definition $\Psi=\langle\Psi\rangle$ returns $h(\langle\Psi\rangle)=0.5$, as it should for a median.)

For the \citet{Hyun2023} radio catalog, an empirical estimate from the \citet{Cotton2018} survey is $\langle\Psi\rangle=0\farcs3$. For $\Psi=0\farcs7$, the FWHM of our A-array observations, $h=0.31$, meaning 69\% of the radio-source population is smaller than 0\farcs7. The radio catalog has  671 such sources in it.  Hence the population is 972 sources, of which 756 are in the catalog, and 216 sources or 29\% of the number in the catalog are predicted to be missing.  Another estimate of $\langle\Psi\rangle$ is that it should match the disk size of {\em resolved} galaxies in the NIRCam images. Measured with NIRCam at 1.15--1.5\,\micron,
$\rm AB<22$\,mag (Figure~\ref{f:fz}) galaxies have a median disk size of 0\farcs25. (See also Figs.~6--8 of \citealt{Windhorst2023}.)  That yields $h=0.27$, a population size of 918 radio sources, and the number missing $\sim$162 sources or 21\% of the sample.

In conclusion, the radio sample is missing at least 10\% and likely less than 29\% of  \citet{Hyun2023}
catalog (6--19 galaxies for the sample within the NIRCam area). Only deeper VLA B- and added C-array images will reveal the
true level of incompleteness.

\section{Basic Data}
\label{sa:data}

\startlongtable
\centerwidetable
\begin{deluxetable*}{rccrrrrrccrrlc}
\tablecaption{Radio counterpart identifications}
\label{t:ID}
%\fontsize{6pt}{7pt}\selectfont
\tabletypesize{\scriptsize}
\tablewidth{0pt}
%\vspace{-3ex}
\tablehead{
\colhead{ID}&
\multicolumn{2}{c}{3\,GHz position} & 
\colhead{$S(3\,GHz$)}&
\multicolumn{2}{c}{F444W position}&
\colhead{sep}&
\colhead{flag}&
\colhead{$a$}&
\colhead{$b$}&
\colhead{PA}&
\colhead{$m(\rm F444W)$}&
\colhead{best $z$}&
\colhead{$Q$}\\
& \colhead{RA}&
\colhead{Decl}&
\colhead{\mmJy} &
\colhead{RA}&
\colhead{Decl}&
\colhead{\arcsec}& 
&
\colhead{\arcsec}&
\colhead{\arcsec}&
\arcdeg&
\colhead{AB}
}
\startdata
193&260.671558&65.710706&21.4&\no&\no&0.94&&\no&\no&\no&\no&\no& \\
194&260.671570&65.711591&81.4&260.671582&65.711598&0.03&em&2.22&0.96&43&17.47&0.1774&4 \\
197&260.674708&65.739125&32.5&260.674714&65.739091&0.12&&0.41&0.20&177&21.84&4.21& \\
201&260.678508&65.768986&50.4&260.678630&65.768949&0.23&e&0.86&0.72&167&19.04&1.1008&4 \\
203&260.679007&65.741979&13.3&260.679013&65.741973&0.02&fm&0.39&0.21&161&22.95&4.38& \\
204&260.679358&65.773721&13.5&260.679405&65.773718&0.07&&0.31&0.23&74&20.44&1.37& \\
213&260.686950&65.722979&27.5&260.686902&65.722975&0.07&fm&1.05&0.74&176&20.41&\no& \\
214&260.687140&65.784347&9&260.687117&65.784353&0.04&p&0.40&0.27&11&20.31&1.22& \\
218&260.690829&65.738099&29.9&260.690855&65.738083&0.07&&0.37&0.32&150&20.08&1.53& \\
219&260.691058&65.751806&15.2&260.691063&65.751772&0.12&s&0.49&0.45&24&19.57&1.22& \\
220&260.691565&65.752944&8.2&260.691558&65.752923&0.07&&0.81&0.36&113&20.51&2.40& \\
223&260.695592&65.784323&13&260.695581&65.784349&0.10&&0.35&0.23&30&20.89&0.82& \\
226&260.697570&65.712623&26.9&260.697592&65.712615&0.04&&0.29&0.22&65&21.07&2.88& \\
232&260.701244&65.776170&50.3&260.701249&65.776179&0.04&&0.31&0.26&24&20.00&1.37& \\
236&260.702588&65.710023&8.3&260.702532&65.710054&0.14&&0.31&0.25&170&20.71&0.82& \\
241&260.708168&65.732152&8.9&260.708113&65.732174&0.12&&0.24&0.21&159&21.16&1.08& \\
244&260.711641&65.809166&5.5&260.711733&65.809144&0.16&ae&0.32&0.14&24&\no&\no& \\
245&260.712947&65.785691&7.1&260.712911&65.785697&0.06&&0.42&0.22&95&21.09&0.88& \\
246&260.714532&65.753392&69&260.714487&65.753354&0.15&s&1.35&0.90&175&18.61&0.5375&4 \\
248&260.715468&65.742532&11.1&260.715284&65.742530&0.27&&0.35&0.30&134&19.99&0.9156&3 \\
251&260.716339&65.711697&22.8&260.716341&65.711693&0.01&&0.47&0.21&35&21.22&1.53& \\
256&260.719510&65.743311&20.2&260.719518&65.743333&0.09&&0.33&0.23&0&20.29&1.45& \\
258&260.719997&65.763175&15.8&260.719987&65.763166&0.03&&0.46&0.36&17&20.37&0.6558&4 \\
260&260.721446&65.813161&57.4&260.721446&65.813168&0.02&e&1.99&1.01&143&18.80&0.5445&4 \\
267&260.725968&65.795907&13.2&260.725955&65.795911&0.02&&0.29&0.17&52&21.85&1.70& \\
268&260.726124&65.812468&18.5&260.726316&65.812485&0.29&om&1.20&0.54&82&18.92&0.31& \\
270&260.728265&65.708595&24.1&260.728243&65.708643&0.18&ae&1.06&0.27&24&\no&\no& \\
271&260.728365&65.801743&17.1&260.728347&65.801763&0.08&&0.49&0.41&157&19.76&0.4558&4 \\
276&260.731090&65.751261&7.5&260.731120&65.751254&0.05&&0.22&0.19&59&23.64&1.08& \\
278&260.731338&65.810081&6.8&260.731473&65.810045&0.24&&0.33&0.19&45&23.36&0.95& \\
280&260.731923&65.730233&11&260.731882&65.730260&0.12&&0.41&0.22&156&20.78&0.88& \\
283&260.733994&65.798503&55.3&260.733997&65.798511&0.03&s&0.69&0.50&25&18.59&0.3278&4 \\
285&260.734793&65.770993&8.9&260.734898&65.770981&0.16&&0.21&0.19&159&22.24&2.08& \\
287&260.736498&65.779211&4.7&\no&\no&\no&&\no&\no&\no&\no&\no& \\
289&260.737916&65.773786&13.5&260.737933&65.773800&0.07&&0.47&0.42&67&21.43&1.98& \\
291&260.739516&65.770729&6.1&260.739377&65.770695&0.23&p&0.74&0.39&94&20.70&0.71& \\
293&260.739960&65.793233&5.4&260.739947&65.793234&0.02&&0.18&0.15&23&24.13&2.75& \\
297&260.745008&65.798769&6.9&260.745009&65.798749&0.06&&0.41&0.29&76&21.04&0.3276&3 \\
298&260.745197&65.792766&12.5&260.745023&65.792808&0.30&&0.34&0.28&50&21.42&\no& \\
300&260.746309&65.784217&18.1&260.746300&65.784221&0.02&&0.23&0.19&23&20.84&2.75& \\
302&260.747308&65.755821&13.6&260.747330&65.755794&0.10&&0.40&0.29&8&20.96&2.08& \\
305&260.748780&65.783504&46.7&260.748774&65.783499&0.02&&0.24&0.19&31&21.38&2.63& \\
306&260.750438&65.770480&17&260.750476&65.770475&0.06&&0.36&0.22&79&21.79&2.88& \\
307&260.750781&65.721062&10.2&260.750750&65.721093&0.12&&0.22&0.17&7&22.16&2.88& \\
308&260.751498&65.735379&8.1&260.751425&65.735380&0.10&s&0.59&0.29&29&19.74&0.1390&4 \\
313&260.755657&65.721894&19.8&260.755584&65.721887&0.10&&0.27&0.18&107&21.51&1.29& \\
314&260.756213&65.713171&324&260.756213&65.713176&0.03&&0.25&0.19&162&20.37&1.53& \\
315&260.757048&65.817354&9.2&260.757018&65.817365&0.06&&0.38&0.33&51&20.97&1.22& \\
319&260.758640&65.800806&103&260.758630&65.800803&0.02&s&0.80&0.57&34&18.11&0.1781&3 \\
325&260.760204&65.822150&8.3&260.760232&65.822166&0.07&&0.24&0.16&6&22.56&2.88& \\
327&260.760675&65.804321&11.7&260.760650&65.804270&0.18&&0.39&0.20&31&20.81&1.22& \\
328&260.761803&65.819778&6.3&260.761849&65.819755&0.10&&0.26&0.21&1&21.69&0.7072&4 \\
329&260.762456&65.815098&5.8&260.762592&65.815153&0.29&&0.34&0.23&119&21.08&0.76& \\
336&260.767063&65.757148&8.1&260.767024&65.757149&0.06&&0.25&0.15&170&22.44&1.70& \\
337&260.767780&65.816263&33.2&260.767742&65.816264&0.06&&0.40&0.23&65&21.16&2.51& \\
338&260.767932&65.779383&19.1&260.767933&65.779387&0.02&&0.32&0.27&174&20.61&1.37& \\
342&260.771767&65.808253&9.9&260.771758&65.808257&0.02&&0.33&0.28&122&22.02&1.62& \\
349&260.776777&65.813659&17.2&260.776796&65.813657&0.03&&0.55&0.35&119&20.24&1.53& \\
355&260.783195&65.801419&8.1&260.783197&65.801389&0.10&&0.25&0.24&64&21.03&1.53& \\
363&260.789215&65.802433&17.1&260.789231&65.802422&0.04&&0.53&0.46&64&19.82&1.5941&3 \\
372&260.793575&65.801777&14.3&260.793510&65.801777&0.09&&0.57&0.45&127&21.11&1.22& \\
373&260.793829&65.795339&19.4&260.793757&65.795353&0.11&&0.80&0.34&154&19.25&0.76& \\
377&260.794800&65.801461&9.6&260.794902&65.801451&0.16&&0.36&0.29&17&21.30&1.37& \\
378&260.794980&65.817308&7.8&260.794968&65.817250&0.20&&0.39&0.21&177&22.49&3.14& \\
\enddata
%\vspace{-3ex}
\tablecomments{$S\rm(3\,GHz)$ is in \mmJy. Source sizes $a$ and $b$ are major- and minor-axis dimensions in arcseconds as determined by SExtractor, and PA is in degrees east from north.  The last column ``Q'' is the quality flag for spectroscopic redshifts.  If missing, redshift is photometric based on available HSC, MMT, and JWST data. The spectroscopic redshift quality code ($Q$) follows the same criteria adopted by the DEEP2 survey \citep{Newman2013}: $Q=4$ indicates a secure redshift, while $Q=3$ indicates a redshift with a 90\% or higher certainty  of being correct.
The flag meanings are as follows:\\
a: the source is outside the LW images but visible in the SW images. The source position, dimensions, and separation from the radio position are from F200W, not the initial automated search on F444W.\\
e: the source is near the image edge, and some flux may be missed.\\
f: the counterpart is a faint source near a brighter one.  The automated search found the brighter object, but the tabulated positions and sizes refer to the correct counterpart.\\
h: spectroscopic redshift from Hectospec.\\
m: position and separation measured manually, but source dimensions are from the automated F444W catalog.\\
o: overlapping sources.\\
p: automated position angle wrong, but effect on photometry should be small.\\
s: photometry aperture may be too small to include a faint outer region of the source.\\}
\end{deluxetable*}

\centerwidetable
\begin{deluxetable*}{rRRRRRRRRRRRrrrl}
\tablecaption{Galaxy properties derived from SED fits}
\label{t:SEDfits}
\tabletypesize{\scriptsize}
\tablewidth{0pt}
%\vspace{-3ex}
\tablehead{
\colhead{ID}&
\colhead{$\chi^2_r$}&
\colhead{\zph}&
\multicolumn{3}{c}{$\log M_*$}&
\multicolumn{3}{c}{$\log \rm SFR$}&
\multicolumn{3}{c}{$A_V$}&
\multicolumn{3}{c}{$f_{\rm AGN}$}&
\colhead{SED class}\\
&&&
\colhead{low}&
\colhead{best}&
\colhead{high}&
\colhead{low}&
\colhead{best}&
\colhead{high}&
\colhead{low}&
\colhead{best}&
\colhead{high}&
\colhead{low}&
\colhead{best}&
\colhead{high}
}
\startdata
194&1.13&0.19&10.73&11.00&11.07&$-$0.65&1.08&0.94&0.31&0.78&0.78&0&0&0.6&RQ Gal \\
197&2.16&4.21&2.40&11.12&10.57&1.07&2.79&2.62&1.55&1.55&1.55&0&0.9&0.9&RQ QSO \\
201&0.62&1.29&10.94&11.46&11.68&0.17&1.46&2.19&0.03&0.78&1.55&0&0.6&0.6&RQ Gal \\
203&0.75&4.38&1.79&10.53&10.72&1.51&1.96&2.02&0.03&0.31&1.55&0&0&0.6&RL QSO \\
204&0.21&1.37&10.50&10.95&11.11&$-$0.65&0.73&1.53&0.03&0.78&1.55&0&0.6&0.6&RL Gal \\
214&0.48&1.22&10.50&11.08&11.15&1.48&1.84&2.20&1.55&1.55&2.33&0&0.3&0.6&RQ Gal \\
218&0.82&1.53&10.84&11.22&11.44&0.09&1.00&1.89&0.31&0.78&1.55&0&0.6&0.6&RL Gal \\
219&0.81&1.22&10.84&11.20&11.40&0.37&1.68&2.18&0.31&0.78&1.55&0&0&0.6&RQ Gal \\
220&1.60&2.40&10.82&11.26&11.40&1.46&2.94&2.62&0.31&1.55&1.55&0&0.6&0.6&RQ QSO \\
223&0.47&0.82&9.77&9.87&10.10&0.69&1.54&1.32&0.31&1.55&1.55&0&0&0.6&RQ QSO \\
226&1.83&2.88&10.92&11.27&11.43&1.19&2.81&2.69&1.55&1.55&1.55&0&0.9&0.9&RQ QSO \\
232&0.27&1.37&9.29&10.92&10.89&1.47&1.57&2.11&1.55&1.55&1.55&0&0.6&0.6&RL QSO \\
236&0.40&0.82&10.02&10.53&10.75&$-$0.76&1.10&1.26&0.31&1.55&1.55&0&0.3&0.6&RQ Gal \\
241&1.61&1.08&9.45&10.52&10.38&$-$1.33&0.93&0.62&1.55&3.10&4.65&0&0.6&0.9&RQ QSO \\
245&0.69&0.88&9.87&10.33&10.54&$-$0.27&1.08&1.39&0.03&0.78&1.55&0&0.3&0.6&RQ Gal \\
246&0.93&0.55&10.79&11.20&11.45&0.10&1.69&1.57&0.03&0.78&2.33&0&0&0.6&RQ Gal \\
248&0.57&0.76&10.50&10.99&11.15&$-$0.35&1.80&1.74&0.39&1.55&1.55&0&0&0.6&RQ Gal \\
251&1.24&1.53&10.22&10.61&10.76&$-$0.16&1.00&1.35&0.03&0.39&1.55&0&0.9&0.9&RL QSO \\
256&1.59&1.45&9.73&10.83&11.01&$-$1.27&0.84&1.04&1.55&2.33&2.33&0&0.9&0.9&RL QSO \\
258&0.67&0.65&10.36&10.46&10.90&$-$0.37&1.31&1.10&0.03&0.78&0.78&0&0&0.6&RQ Gal \\
260&0.36&0.60&11.08&11.43&11.48&$-$0.59&$-$0.46&1.10&0.03&0.78&1.55&0&0&0.6&RL Gal \\
267&2.26&1.70&10.13&10.38&10.71&$-$0.86&1.03&1.71&1.55&1.55&2.33&0&0.9&0.9&RL QSO \\
268&0.54&0.31&10.42&10.87&10.91&$-$0.82&0.67&0.75&0.03&0.78&1.55&0&0&0.6&RQ Gal \\
271&0.92&0.31&10.22&10.26&10.88&$-$4.81&0.91&0.38&0.03&0.78&0.78&0&0&0.9&RQ Gal \\
276&1.45&1.08&8.56&8.98&9.53&$-$1.12&0.23&0.22&0.03&0.78&1.55&0&0&0.6&RL Gal \\
278&2.02&0.95&2.21&8.68&8.78&$-$1.57&0.21&$-$0.14&1.55&1.55&2.33&0&0.3&0.6&RL QSO \\
280&0.53&0.88&9.87&10.35&10.57&$-$0.81&1.04&1.36&0.31&1.55&1.55&0&0.3&0.6&RQ Gal \\
283&0.65&0.31&10.53&11.03&11.06&$-$0.63&0.86&0.99&0.03&0.31&0.78&0&0.3&0.9&RQ Gal \\
285&1.63&2.08&10.18&10.81&10.81&$-$0.23&$-$0.40&1.74&0.03&1.55&1.55&0&0.9&0.6&RL Gal \\
289&0.76&1.98&10.65&11.11&11.10&$-$0.14&1.06&1.96&0.03&0.78&1.55&0&0.9&0.9&RL Gal \\
291&0.46&0.71&10.18&10.59&10.79&$-$3.12&0.18&0.34&0.03&0.78&1.55&0&0&0.6&RQ Gal \\
293&2.04&2.75&1.67&9.71&9.80&$-$0.09&1.37&1.27&0.03&0.39&1.55&0.3&0.9&0.9&RQ QSO \\
297&0.49&0.31&9.58&9.74&10.18&$-$1.55&0.32&0.12&0.02&0.31&0.39&0&0&0.6&RQ Gal \\
298&0.09&1.37&10.01&10.59&10.68&0.07&0.99&1.63&0.03&0.78&1.55&0&0.3&0.6&RQ Gal \\
300&1.49&2.75&10.25&10.90&11.38&0.71&2.56&2.64&1.55&1.55&1.55&0&0.6&0.6&RQ Gal \\
302&1.88&2.08&10.22&10.81&11.17&1.79&1.91&2.10&0.31&0.78&0.78&0&0.9&0.9&RQ QSO \\
305&2.35&2.63&10.20&11.02&11.17&0.45&1.73&2.20&0.03&1.55&1.55&0&0.9&0.6&RL QSO \\
306&0.29&2.88&2.46&10.73&10.27&0.44&1.98&1.93&0.31&0.78&0.78&0&0.9&0.9&RQ QSO \\
307&1.74&2.88&10.46&10.62&11.05&0.26&1.31&2.14&0.39&1.55&1.55&0&0.9&0.6&RL Gal \\
308&0.63&0.11&9.77&10.00&10.09&$-$1.81&$-$0.17&$-$0.16&0.02&0.02&0.39&0&0&0.6&RQ Gal \\
313&1.64&1.29&10.12&10.37&10.65&$-$0.68&0.38&1.55&0.39&1.55&1.55&0&0.6&0.6&RL Gal \\
314&1.00&1.53&10.46&10.70&11.06&$-$0.24&1.11&1.95&0.31&0.78&1.55&0.3&0.9&0.9&RL QSO \\
315&2.22&1.22&10.36&11.03&10.99&$-$1.63&$-$1.77&0.68&0.39&2.33&2.33&0&0.9&0.9&RL QSO \\
319&1.07&0.19&10.17&10.42&10.56&1.01&1.18&1.35&0.03&0.03&0.78&0&0.3&0.6&RQ Gal \\
325&1.97&2.88&10.32&10.78&11.05&$-$0.28&1.19&1.99&1.55&2.33&3.10&0&0.9&0.9&RL QSO \\
327&0.42&1.22&10.16&10.42&10.72&$-$0.23&2.09&1.79&0.31&1.55&1.55&0&0&0.6&RQ Gal \\
328&0.47&0.65&9.51&10.02&10.30&$-$0.67&0.81&0.80&0.03&0.31&0.78&0&0&0.6&RQ Gal \\
329&0.31&0.76&10.03&10.44&10.69&$-$3.38&$-$1.39&0.03&0.03&1.55&1.55&0&0&0.6&RL Gal \\
336&1.76&1.70&9.77&10.36&10.44&$-$0.17&0.36&1.48&0.03&0.39&1.55&0&0.9&0.6&RL Gal \\
337&1.97&2.51&10.66&10.87&11.24&0.24&2.53&2.49&1.55&2.33&2.33&0&0&0.6&RQ Gal \\
338&0.22&1.37&10.44&10.91&11.02&0.25&1.98&1.99&0.03&1.55&1.55&0&0&0.6&RQ Gal \\
342&2.04&1.62&9.86&10.37&10.55&$-$0.75&$-$0.85&1.05&0.39&1.55&2.33&0&0.9&0.6&RL Gal \\
349&1.19&1.53&10.47&11.26&11.15&0.56&1.67&2.43&0.39&1.55&1.55&0&0&0.6&RQ Gal \\
355&0.78&1.53&10.35&10.50&10.80&0.73&2.04&2.01&0.03&0.78&0.78&0&0.3&0.6&RQ QSO \\
363&1.36&3.01&10.54&11.19&11.30&1.70&2.00&2.19&0.03&0.78&1.55&0.3&0.3&0.3&RQ QSO \\
372&0.56&1.22&10.15&10.42&10.69&0.06&1.27&1.73&0.03&0.78&1.55&0&0.6&0.6&RQ Gal \\
373&0.31&0.76&10.49&11.12&11.28&$-$0.65&1.53&1.67&0.39&1.55&1.55&0&0&0.6&RQ Gal \\
377&1.03&1.37&10.19&10.23&10.81&$-$0.10&1.90&1.72&0.31&1.55&1.55&0&0.3&0.6&RQ Gal \\
378&0.33&3.14&1.65&10.50&10.56&1.12&2.03&1.95&0.03&0.39&0.78&0&0.9&0.3&RQ QSO \\
\enddata
\tablecomments{Data are based on fitting the HSC, MMT, and JWST data with CIGALE.  Columns labeled ``low'' are parameter values at the 16th percentile of the PDF for that parameter.  Columns labeled ``best'' are parameter values for the model with the lowest $\chi^2$.  Columns labeled ``high'' are parameter values at the 84th percentile of the PDF.  Column 3 gives the best-fit photometric redshifts, but when a spectroscopic redshift is available, other parameters are shown for the redshift fixed at the spectroscopic value (Table~\ref{t:ID}).  For ID~363, the photometric redshift shown in the table is a catastrophic failure.}
\end{deluxetable*}

\end{document}